\documentclass[fleqn,12pt,a4paper]{article}
\usepackage{amsmath}
\usepackage{graphicx}
\usepackage{txfonts}
\begin{document}

\rightline{KANAZAWA-19-01}
\vskip15mm
\begin{center}
{\LARGE
Logical Reasoning \\
\vskip1mm
for Revealing the Critical Temperature\\
\vskip1mm
through Deep Learning of Configuration Ensemble\\
\vskip2mm
of Statistical Systems}
\end{center}

\vskip5mm
\centerline{Ken-Ichi Aoki\footnote{aoki@hep.s.kanazawa-u.ac.jp}, 
Tatsuhiro Fujita\footnote{t\_fujita@hep.s.kanazawa-u.ac.jp},
and Tamao Kobayashi$^\dagger$\footnote{kobayasi@yonago-k.ac.jp}
}

\centerline{Institute for Theoretical Physics, Kanazawa University, 
Kakuma-machi, Kanazawa 920-1192, Japan}
\centerline{$^\dagger$ Yonago College, National Institute of Technology, 
4448 Hikona-machi, Yonago 683-8502, Japan}
\vskip10mm
\centerline{
[Published in Journal of the Physical Society of Japan {\bf 88}, 054002 (2019)]}

\vskip10mm
\noindent {\Large\bf Abstract}

Recently, there have been many works on the deep learning of statistical ensembles 
to determine the critical temperature of a possible phase transition. 
We analyze the detailed structure of an optimized deep learning machine and 
prove the basic equalities among the optimized machine parameters and the
physical quantities of the statistical system. 
According to these equalities, we conclude that the bias
parameters of the final full connection layer record the free energy of the
statistical system as a function of temperature.
We confirm these equalities in 
one- and two-dimensional Ising spin models and actually demonstrate that the deep learning
machine reveals the critical temperature of the phase transition through the
second difference of bias parameters, which is equivalent to the specific heat.
Our results disprove the previous works claiming
that the weight parameters of the full connection might play a role of the order
parameter such as the spin expectation.

\def\Beq{\begin{equation}}
\def\Eeq{\end{equation}}
\def\Bea{\begin{eqnarray}}
\def\Eea{\end{eqnarray}}
\def\Haii{\{\sigma\}}

\newcommand{\argmax}{\mathop{\rm arg~max}\limits}
\newcommand{\argmin}{\mathop{\rm arg~min}\limits}
\newcommand{\softmax}{\mathop{\rm softmax}\limits}
\newcommand{\average}{\mathop{\rm average}\limits}

\section{Introduction and Summary}

Recently, deep learning has been drawing attention
in various applications of image recognition or artificial intelligence technology, 
since it has been very successful beyond expectation. 
In other words, why deep learning is so effective in these fields
has not yet been clarified.

On the other hand, 
the renormalization group\cite{Wilson1,Wilson2,Wilson,Kadanoff} in physics has many common
characteristics with the deep learning. The input data for the 
deep learning corresponds to the microvariables in physics, and
the convolution filtering resembles the renormalization 
transformation itself, which picks up the main structural 
features of the input data, that is, the relevant operators
(interactions) in the renormalization group terminology.\cite{PD,KT,iso,SL,SJYJ,M.Koch}

Therefore, analyzing these similarities between the deep
learning and the renormalization group would have twofold benefits: clarifying the origin and mechanism of 
the effectiveness of deep learning will shed light on 
the strategies for improving the deep learning system
and for developing new types of renormalization group.

In this work, we focus on the recent interest in the fact 
that the deep learning of the configuration ensemble of a statistical system
may indicate the phase transition temperature of the system through
the optimized machine parameters, which may behave as the order
parameter of the system.\cite{tanaka_tomiya,ohzeki}
However, there has been no logical 
argument or understanding of why the optimized machine parameter
can possibly behave as the order parameter showing the phase transition
temperature.
Therefore, we analyze a deep learning machine in detail and will clarify 
what is learned by the machine and how it learns.

\begin{figure}[!h]
\vskip7mm
\begin{center}
\includegraphics[clip,scale=0.14]{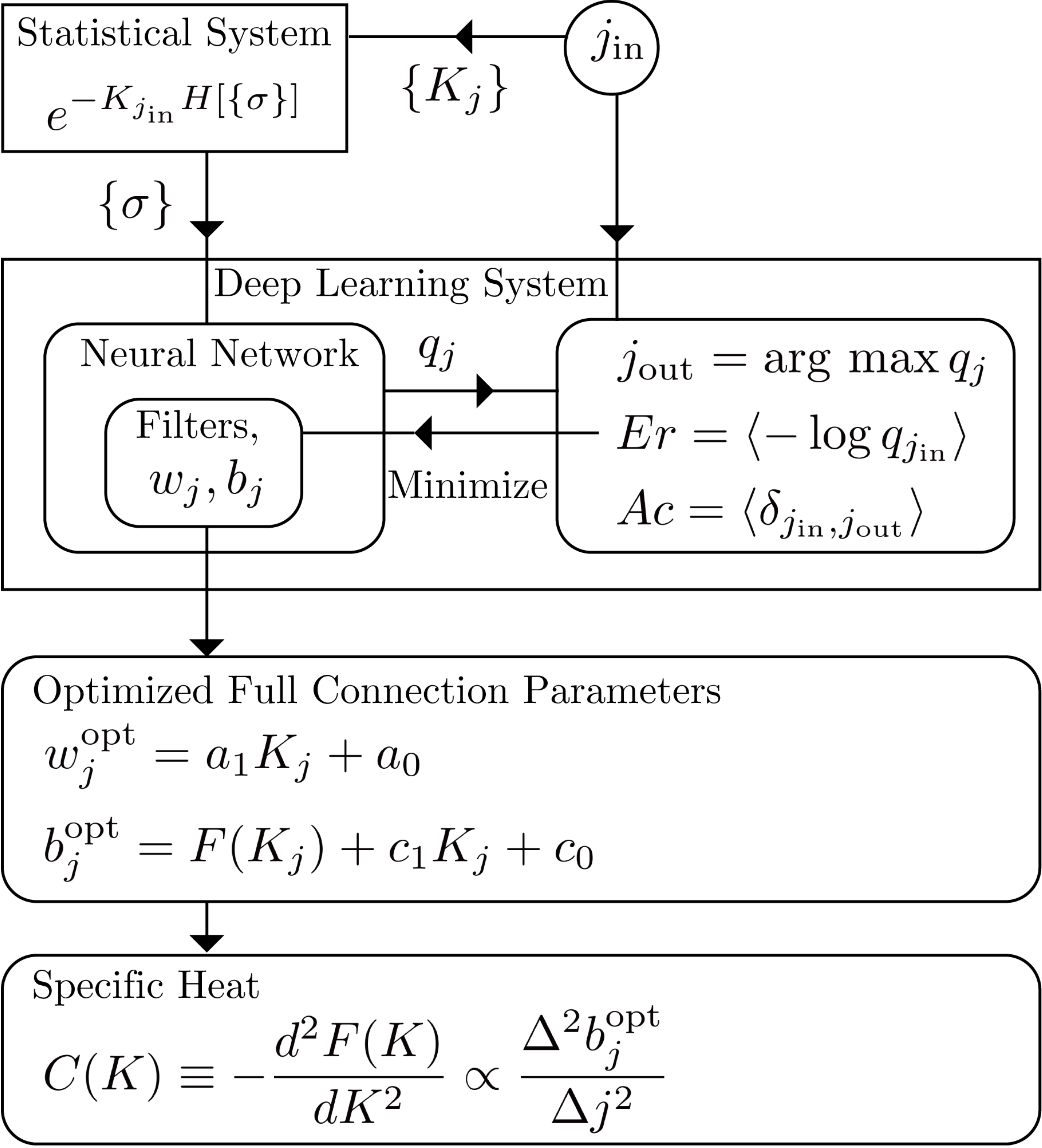}
\caption{What does deep learning learn
from statistical system configurations?}
\label{summary}
\end{center}
\vskip-8mm
\end{figure}

\begin{figure}[!h]
\vskip7mm
\begin{center}
\centerline{\includegraphics[clip,scale=0.12]{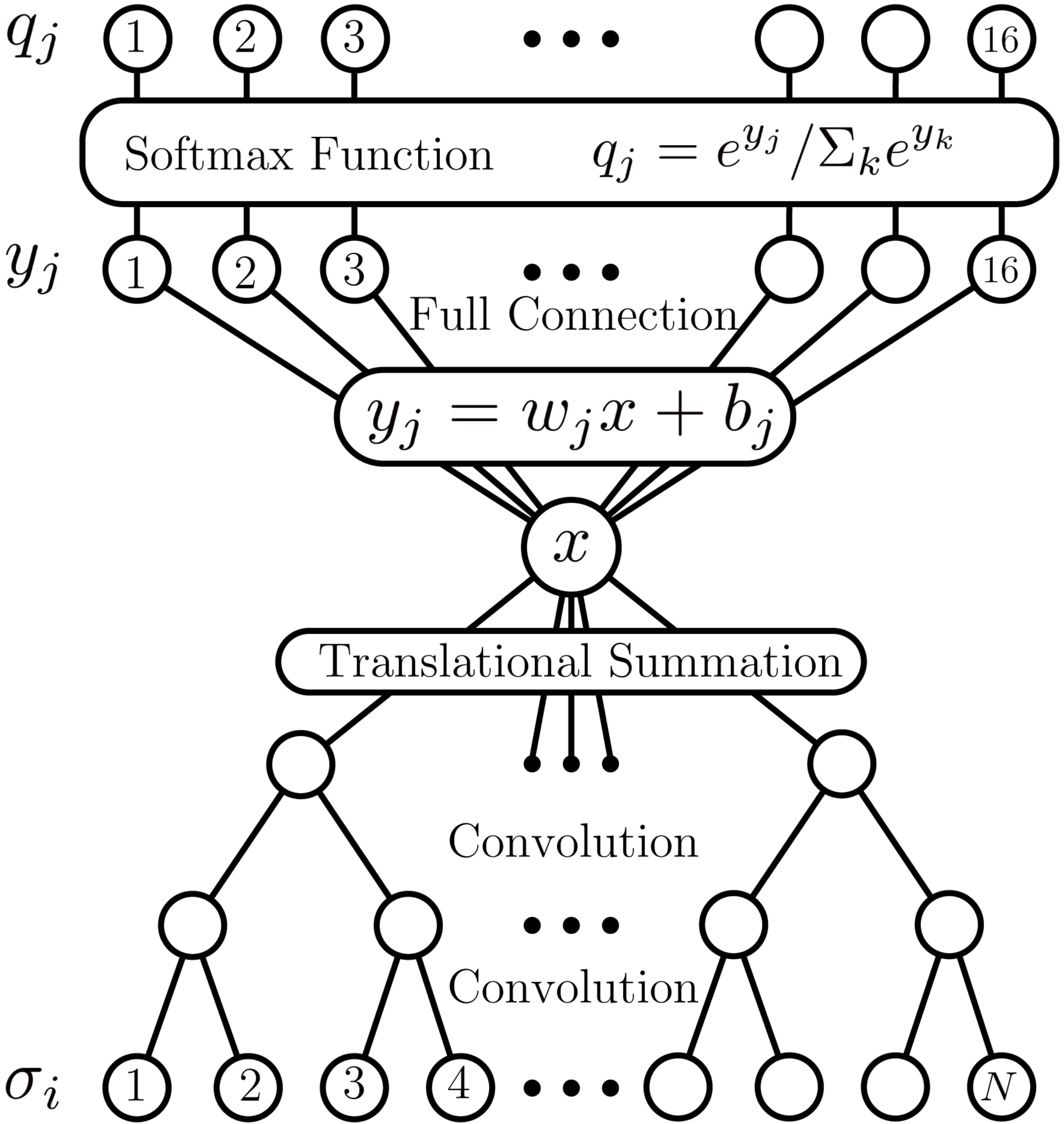}}
\caption{Deep learning machine structure.}
\label{machinestructure}
\end{center}
\vskip-8mm
\end{figure}

Here, we summarize our main results referring to Fig.~\ref{summary}, which
explains the logical flow of 
the machine learning, its optimized parameters, and
the information of physical quantities to be obtained.
The machine structure is drawn schematically in Fig.\ref{machinestructure}.

We prepare a statistical system where the physical 
microvariables are the spin $\sigma$, and 
$\{\sigma\}$ denotes a configuration of the system.
The statistical weight of a configuration $\{\sigma\}$ is
proportional to 
\Beq
\exp \left[ -K H(\{\sigma \}) \right],
\label{statisticalweight}
\Eeq
where $H$ is the Hamiltonian of the system and $K$ is the 
(inverse) temperature.

We pick up an appropriate set of temperatures
$K_j, j=1,2,\cdots 16$, where we take 16 values
in this article. Taking one temperature, we make up a
system configuration and transfer it
to the machine with a label of temperature number $j_{\rm in}$.
This type of learning is called supervised learning. 
Note that only the temperature number $j_{\rm in}$ labels the 
configuration data. The actual value of $K_{j_{\rm in}}$ does not matter
at all. Furthermore, as will be clarified later, 
these temperatures do not have to be equally spaced 
or even ordered (e.g., increasing order). 

The machine receives the configuration with label $j_{\rm in}$ as 
input data, which is shown in the lowest layer
in Fig.~\ref{machinestructure}.
The target of the machine is to find the temperature number that 
generates the configuration. Of course, there cannot be a unique
answer for the temperature of a configuration. In fact, any configuration
can occur at any temperature, ignoring its actual probability of occurrence.
Therefore, the machine should learn the best answer for any configuration, 
that is, the probabilistic correctness of the answer should be maximized.

The machine is constructed according to the standard multilayered neural 
network system. First, we set convolution layers, which multiply the lower layer variables by filters to make the upper layer variables.
This is similar to the renormalization group transformation.
Actually, we work with multiple channel filters but they are not shown in
Fig.~\ref{machinestructure}.

After a number of convolution layers, we average all the 
variables to define only one variable $x$. 
This averaging policy is based on the
translational invariance in the original space direction 
of our statistical system.
Since the configurations are totally translationally invariant,
the optimized machine parameters must be translationally invariant.
When working with multichannel convolutional layers, they are 
finally summed up to make the single variable $x$.

This intermediate variable $x$ is of essential importance.
This single variable must contain sufficient information to 
best predict the input temperature.
As proved in the next section, $x$ must be the Hamiltonian 
of the system (up to the normalization and origin).

Then, we set the full connection layer from $x$ to $y_j, j=1,2,\cdots 16$,
through linear functions,
\Beq
y_j = w_j x+ b_j,
\label{defofyj}
\Eeq
where $w_j$ is the weight and $b_j$ is the bias for each output channel.

Finally, we apply the softmax function to $y_j$ and obtain the machine
output $q_j$,
\Beq
q_j = \frac{e^{y_j}}{\Sigma_k e^{y_k}}.
\label{defofqj}
\Eeq
We interpret that this $q_j$ is the posterior probability of temperature
$K_j$ determined by the machine for the input configuration $\{\sigma\}$.

As shown in Fig.~\ref{summary}, we define $j_{\rm out}$ by the maximum
valued $q_j$, which should be understood as the predicted temperature.
If the prediction $j_{\rm out}$ is equal to $j_{\rm in}$, then the case 
is judged as a correct answer, which totally defines the accuracy of the
machine,
\Beq
Ac=\langle \delta_{j_{\rm in}, j_{\rm out}}\rangle,
\label{accuracyfunction}
\Eeq
where $\langle\cdot\rangle$ represents the expectation value
for the whole input ensembles of mixed-temperature.
The machine parameters, convolution filters, and full connection parameters
are optimized to lower the standard error function
\Beq
Er=\langle -\log q_{j_{\rm in}}\rangle,
\label{errorfunction}
\Eeq
by using the stochastic gradient descent.

The level of optimization can be evaluated by comparing the machine
achievement with the theoretical bounds. In the models treated in this article,  
we can calculate the theoretical upper limit of the accuracy 
and the theoretical lower limit of the error with sufficient precision.
After the numerous learnings (optimization), 
we retrieve the full connection 
parameters $w_j$ and $b_j$. 

By a simple and logical argument presented in Sec.~4,
we obtain the relations
that these optimized machine parameters should obey.
After simple manipulation, we obtain the general solution for 
these optimized parameters as written in Fig.~\ref{summary}:
\Bea
w_j^{\rm opt}&=&a_1 K_j + a_0, \label{wopt}\\
b_j^{\rm opt}&=&F(K_j) + c_1 K_j + c_0, \label{bopt}
\Eea
where $a_1, a_0, c_1,$ and $c_0$ are $j$-independent arbitrary constants. 
The weight $w_j$ is a linear function in $K_j$, and the
bias $b_j$ is the free energy $F$ of the original statistical system
up to an arbitrary  linear function in $K_j$.
In our notation, the free energy is defined as
\Beq
F(K_j) \equiv - \log Z(K_j), 
\Eeq
and it is a dimensionless quantity, 

These relations are remarkable, which were reported first by 
the authors.\cite{AFK2018}
They relate the optimized machine parameters after 
learning with the physical quantities of the input statistical system.
The first equation Eq.~(\ref{wopt}) gives us the actual temperature values
up to the unit normalization ($a_1$) and the zero origin of the
inverse temperature ($a_0$).

The second equation Eq.~(\ref{bopt}) declares that the bias is simply the free energy
of the statistical system at temperature $K_j$ up to the total
normalization of the statistical weight ($c_0$) 
and the zero origin of the Hamiltonian ($c_1$).

Note that the above emergence of the arbitrary constants $a_1, c_0,$  and $c_1$ is
inevitable, and they cannot be determined, that is, they have
no physical significance here.
The parameter $a_0$ is in general physically meaningful,
but in this story of
deep learning determination, it becomes irrelevant.

With these relations, we can calculate the second derivative 
of the free energy with respect to the temperature.
The temperature spacing and order can be determined 
using Eq.~(\ref{wopt}). Then, the second derivative of
free energy, that is, the specific heat, can be evaluated 
using the second-order difference of $b_j$, without being 
disturbed by the arbitrary parameters $a_1, a_0, c_1,$ and $c_0$.

Now, the phase transition, if any and if enforcing singular
peak behavior of the specific heat, is revealed through
the second difference of the bias $b_j$, and the critical
temperature number can be read off. 
In the rest of this section, 
we apply the above summarized strategy to the 
two-dimensional nearest-neighbor Ising model (2d-NNI) and 
present the concluding plots. 

\begin{figure}[!h]
\begin{center}
\includegraphics[clip,scale=0.6]{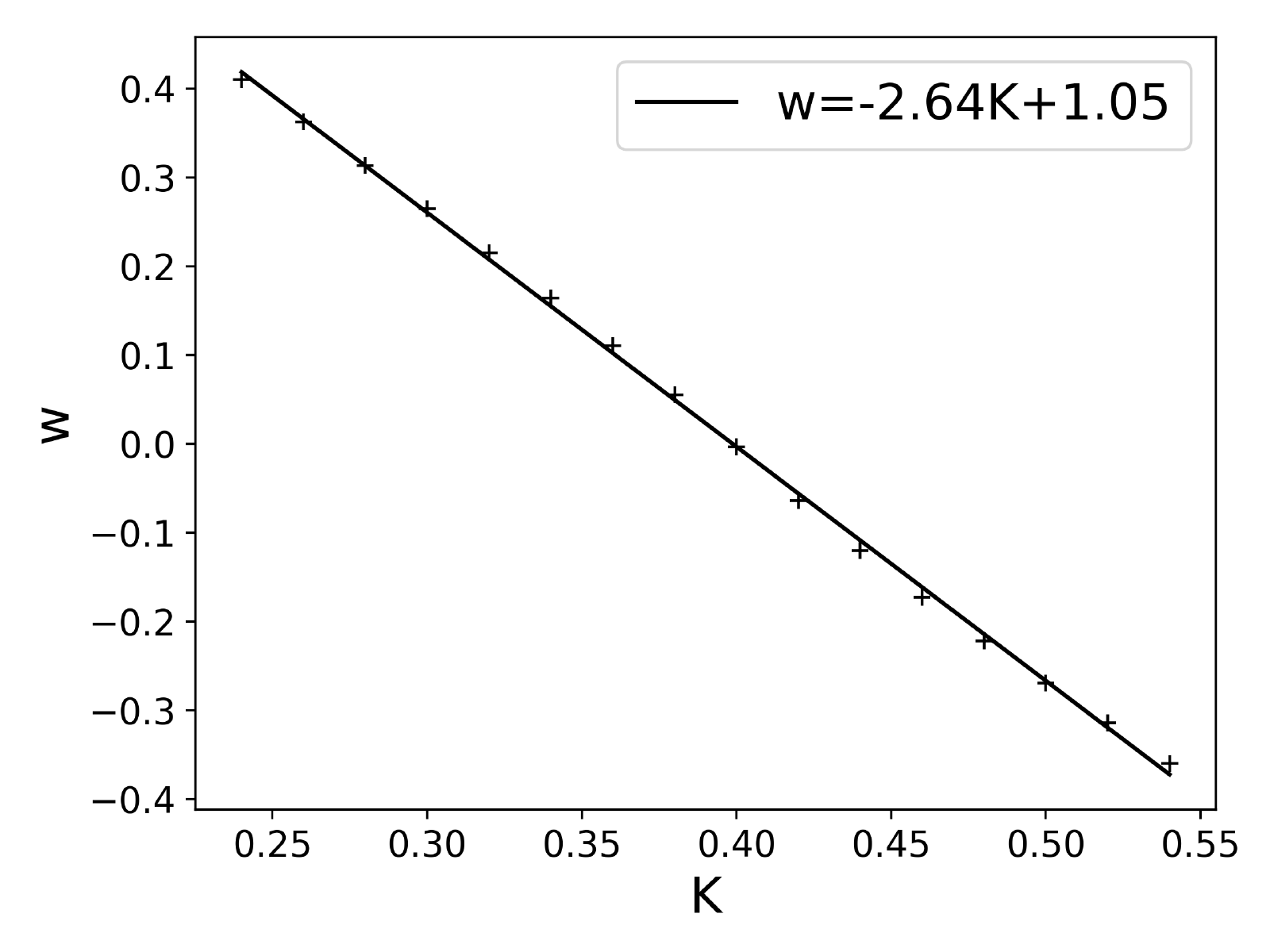}
\caption{Linear dependence of optimized weight $w_j$ on $K$ 
(2d-NNI).}
\label{two-Kw}
\end{center}
\end{figure}

We work with a $32\times 32$-size square lattice. 
As the input temperatures, we take 16 values in 
$[0.24, 0.54]$ with equal spacing (0.02) for simplicity.
We plot 
the optimized full connection weight 
parameters vs $j$ or $K_j$ in Fig.~\ref{two-Kw}, which 
clearly shows that
$w_j$ is very well fitted by a linear function in $K_j$.

\begin{figure}[!h]
\begin{center}
\includegraphics[clip,scale=0.6]{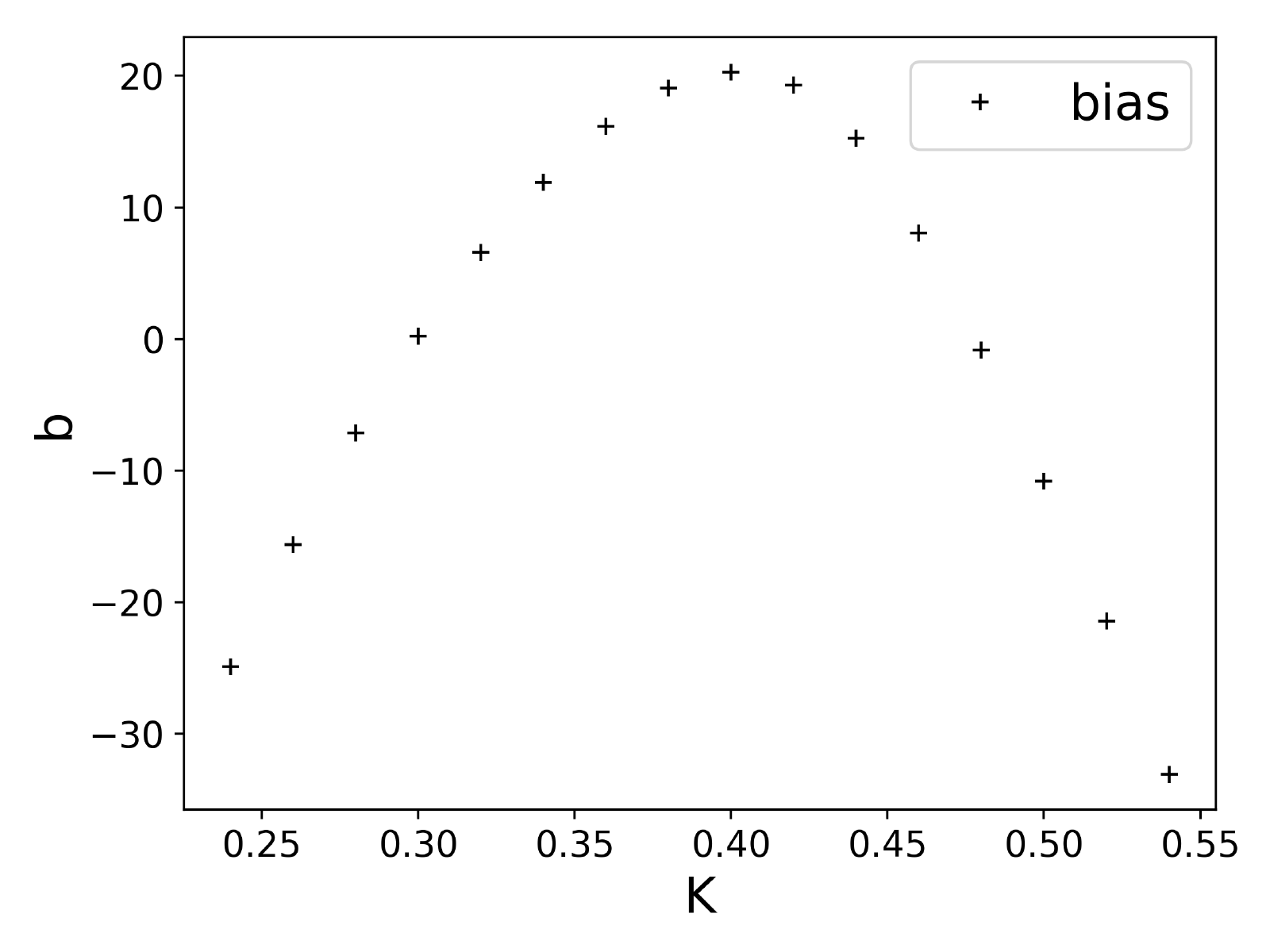}
\caption{Optimized bias $b_j$  (2d-NNI).}
\label{two-Kb}
\end{center}
\end{figure}

Figure~\ref{two-Kb} plots the optimized full connection bias
parameters vs $j$ or $K_j$. It shows that the bias is a function 
of $K_j$ with the second derivative.

\begin{figure}[!h]
\begin{center}
\includegraphics[clip,scale=0.6]{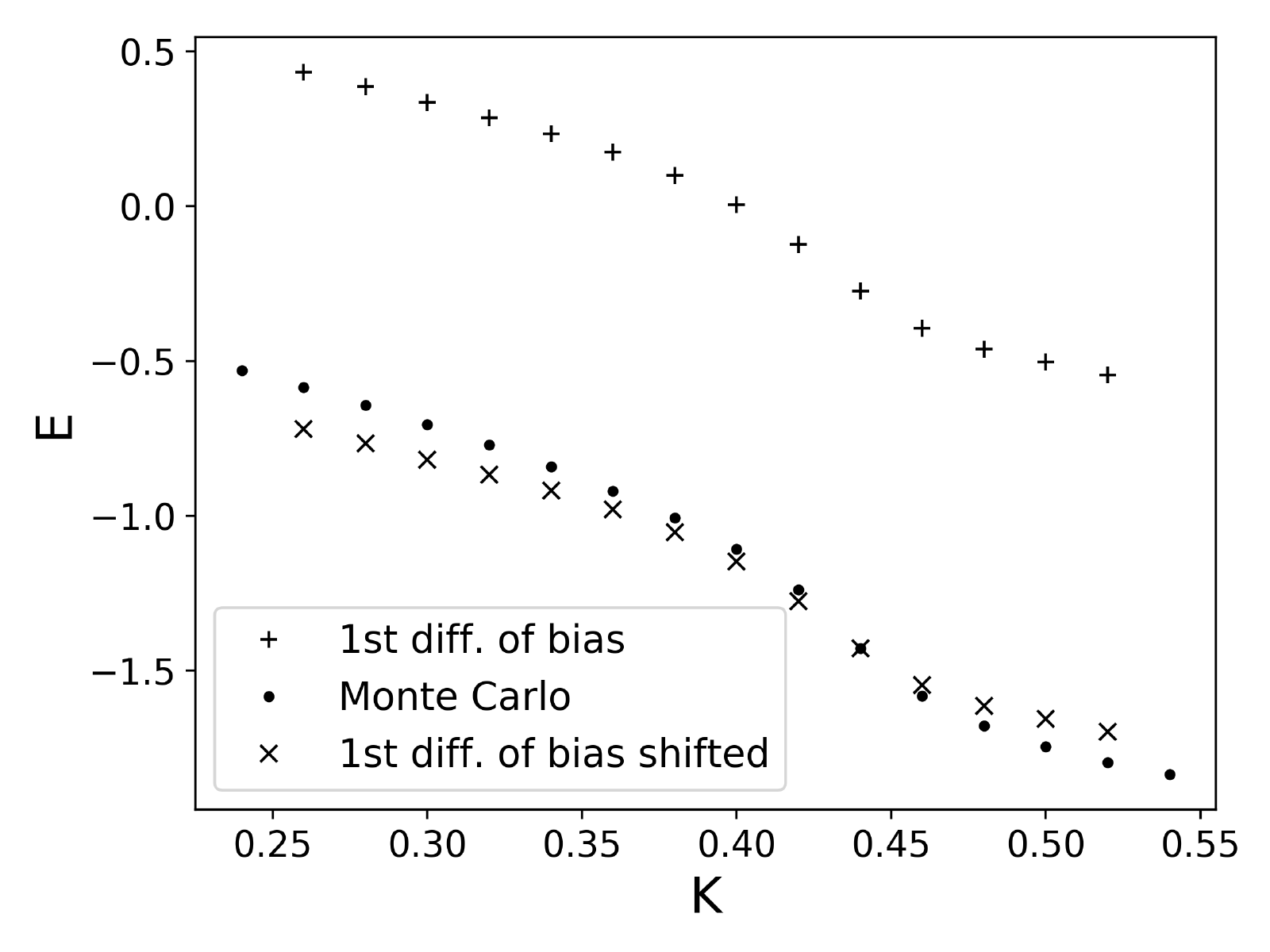}
\caption{First difference of bias and energy 
expectation given by Monte Carlo simulation (2d-NNI).}
\label{two-E}
\end{center}
\end{figure}

To evaluate the physical 
quantities, we take the first difference of $b_j$, which 
corresponds to the energy expectation value $E$ at 
$K_j$. In Fig.~\ref{two-E}, we plot the first difference of $b_j$ compared with
the exact energy expectation values obtained 
by Monte Carlo (MC) simulation.
Also, we plot a shifted first difference of $b_j$ so that it is equal to the
MC simulation value at $K=0.44$. 
Note that the origin of the Hamiltonian has no significance here as
$c_1$ is arbitrary.
The shifted first difference of $b_j$ well approximates the exact 
energy expectation values.

\begin{figure}[!h]
\begin{center}
\includegraphics[clip,scale=0.6]{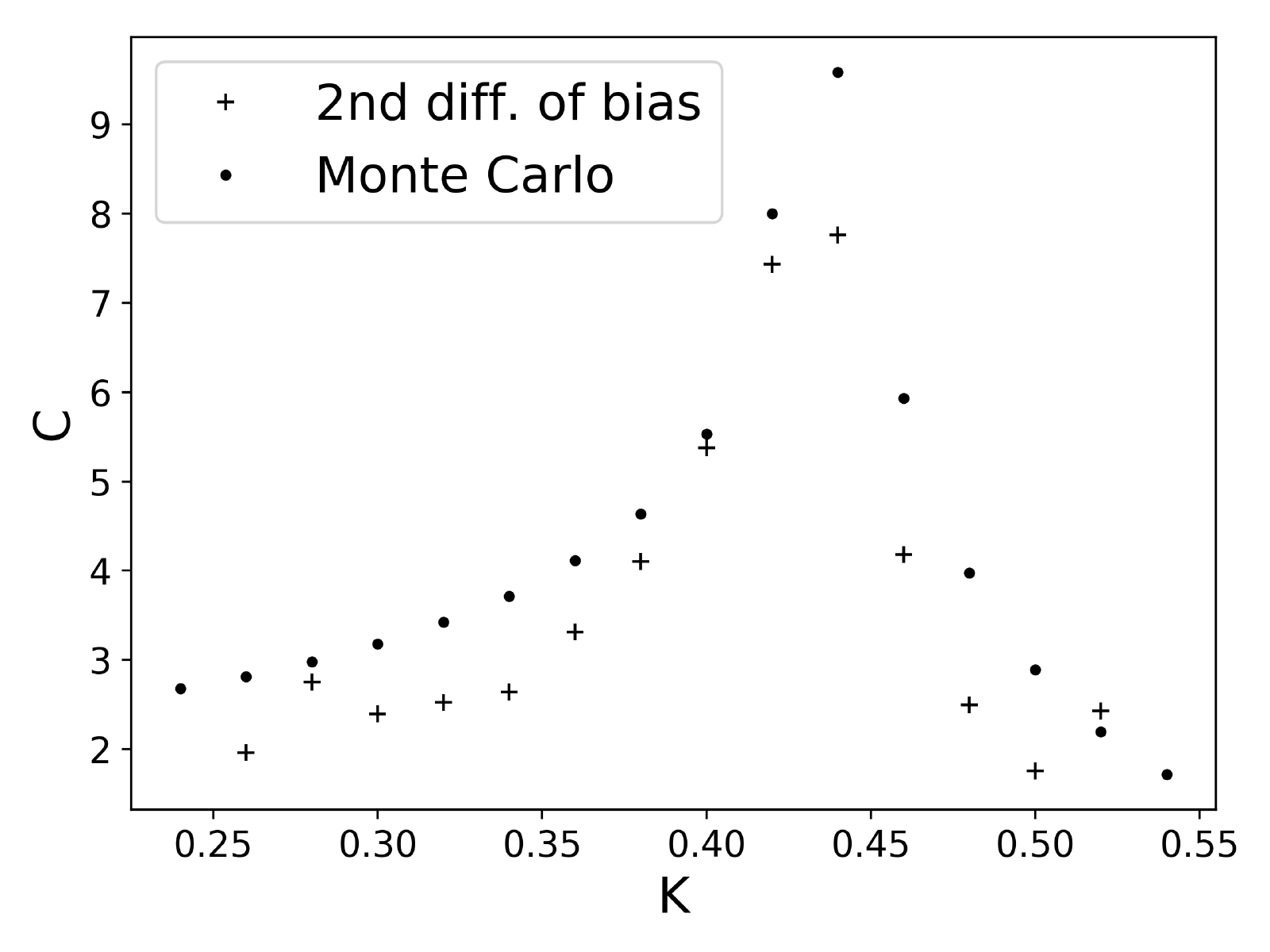}
\caption{Second difference of bias and specific heat 
given by Monte Carlo simulation (2d-NNI).}
\label{two-C}
\end{center}
\end{figure}

Then, we proceed to the second difference of the bias
parameter, which corresponds to the specific heat.
In Fig.~\ref{two-C}, 
we compare the results with exact specific heat 
values given by MC simulation.
Owing to the finiteness of our system, the specific heat does 
not diverge.
We see the characteristic increase in the specific heat
near the phase transition point $K_{\rm c} \simeq 0.44$
(in case of the infinite volume limit).\cite{Onsager}
Also, note that there are no remaining arbitrary constants,
and the second difference of the bias quantitatively approximates 
the exact specific heat well.
Thus, the specific heat singularity is in fact
engraved in the optimized bias parameters.

These results are completely different from in statements
in previously published materials\cite{tanaka_tomiya,ohzeki}, that is,  
the optimized weight parameter works as an
order parameter such as the spin expectation value as a function of temperature.
However, these previous statements were just
claimed without any logical or even plausible argument.
As discussed in the next section, 
the single unique variable $x$ in the machine must represent
the Hamiltonian as long as the machine is tuned to be optimistic.
Then, the full connection 
weight parameter $w_j$, which becomes a dimensionless
quantity by being multiplied by $x$, must have the dimensionality of
the inverse energy. Actually in our solution, the weight $w_j$
is a linear function of $K_j$ and satisfies the correct dimensionality.
Thus, $w_j$ cannot represent the spin expectation value,
which is concluded even from this dimensionality argument only.

\section{Optimized Machine Must Know Hamiltonian}

In this section, we prove that the optimized machine
must know the Hamiltonian of the system as a function of
the configuration.
The statistical system is defined by the statistical weight given in
Eq.~(\ref{statisticalweight}). In general, the target external control
parameter is multiplied by the conjugate physical quantity in the 
exponent.
Here, the target parameter is the inverse temperature $K$ and the
conjugate is called the energy or the Hamiltonian $H(\Haii)$.

The normalized probability of occurrence of a configuration $\Haii$ is given by
\Beq
P(\{\sigma\};K) = \frac{\exp(-K H(\{\sigma\}))}{Z(K)},
\label{probabilityofhaii}
\Eeq
where the partition function $Z(K)$ is defined by summing up 
all possible configurations,
\Beq
Z(K)=\sum_{\{\sigma\}} \exp(-K H(\{\sigma\})).
\Eeq

We set a number of different values for the inverse temperature 
$K_j, j=1,2,\cdots,J$. In this article, we take $J=16$.
The total ensemble is the mixed set of configurations of the 
$J$-sort of different temperatures. In this mixed-temperature ensemble, 
the occurrence probability of $\Haii$ is
\Beq
P(\{\sigma\})= \frac{1}{J} \sum_j P(\Haii;K_j)
 = \frac{1}{J}\sum_j\frac{\exp(-K_j H(\{\sigma\}))}{Z(K_j)},
\Eeq
where we fix the number of configurations for each 
temperature to be equal.

Suppose we find a configuration $\Haii$, then the posterior
(conditional) probability of 
temperature number $j$ is given by
\Beq
Q(j;\Haii)=\frac{P(\Haii;K_j)}{\sum_k P(\Haii;K_k)}.
\label{posteriorinput}
\Eeq
Consider the machine output for the input configuration $\Haii$.
If the machine outputs the temperature number $j$, its
accuracy, the rate of successful guesses, is simply 
this $Q(j;\Haii)$.

Therefore, to maximize the accuracy, the machine should output 
$j_{\rm max}$ that maximizes $Q(j;\Haii)$,
\Beq
j_{\rm max}(\Haii) = \argmax_j Q(j;\Haii).
\label{argmaxaccuracy}
\Eeq
This $j_{\rm max}$ is a function of $\Haii$ and the maximum
accuracy $A$ is also a function of $\Haii$,
\Beq
A(\Haii) = \max_j Q(j;\Haii) = Q(j_{\rm max};\Haii) .
\Eeq
By averaging this with the whole input $\Haii$, we have the maximum value of
the accuracy for the total ensemble,
\Bea
A_{\rm max}\kern-4mm&=&\kern -4mm \langle A(\Haii) \rangle \nonumber \\
&=&
\sum_{\Haii}\frac{1}{J} \left[\sum_j \frac{\exp(-K_j H(\{\sigma\}))}{Z(K_j)}\right]
\max_i \left[\frac{P(\Haii;K_i)}{\sum_k P(\Haii;K_k)}\right].
\label{defoftotalaccuracy}
\Eea

The maximum accuracy $A(\Haii) $ and the maximum guess 
$j_{\rm max}(\Haii)$ are both functions of $\Haii$. 
It is important that their 
dependence comes only through $P(\Haii;K)$, that is, through
the Hamiltonian $H(\Haii)$.
We can denote this restricted dependence explicitly as 
\Bea
A(\Haii) \Rightarrow A(H(\Haii)), \\
j_{\rm max}(\Haii) \Rightarrow j_{\rm max}(H(\Haii)).
\Eea
That is, $A$ and $j_{\rm max}$ are functions of a single variable:
the Hamiltonian of $\Haii$.

Now, we understand that the optimized machine 
to achieve the theoretical upper bound of the total accuracy must
be able to correctly calculate the Hamiltonian as a function of $\Haii$.
Also, there is no way of improving the accuracy by adding information about
quantities other than the Hamiltonian. Although 
not literally correct, we would like to express this situation as follows: knowing 
the Hamiltonian is a {\sl necessary and sufficient} condition for the machine 
to accomplish the highest limit of accuracy.
 
Here, we should note a subtle point.
The deep learning machine is actually optimized by the 
condition of minimizing the error function in Eq.~(\ref{errorfunction}), 
not by maximizing the accuracy function in Eq.~(\ref{accuracyfunction}).
Minimizing the error function is a much stronger condition than maximizing 
the accuracy. 
This fact can be understood quickly by looking at Fig.~\ref{yfunction}, which will be
explained later in Sec.~8.
In fact, the minimized error machine necessarily achieves the highest 
accuracy, but the reverse does not  hold.
In this sense, the above statement of the necessary and sufficient 
condition of knowing the Hamiltonian should apply rather to the minimum
error machine. However, if we prepare an initial set of temperatures 
so that the spacing between neighboring temperatures is sufficiently small, then 
the difference between these two types of machines will decrease and
finally vanish in the limit of infinitesimal temperature discrimination.

We extend the above argument to the case where there exist other
external control parameters in the statistical weight.
For example, we consider the uniform external magnetic field $f$,
which interacts with the total spin of the system 
$S(\Haii)$. Then, the statistical weight reads
\Beq
\exp (-K H(\{\sigma \}) + f S(\Haii)).
\Eeq
In fact, this statistical weight is unphysical since the interactions 
between the spin and the external magnetic field should be included in the
Hamiltonian of the system coupled with the temperature. 
Of course, in the case that the Hamiltonian contains an external field
parameter, there is no change in the previous conclusion, that is, 
the Hamiltonian function including external field interactions 
must be learned by the machine.

However, as a purely theoretical 
system, we can make an ensemble of the above type of system.
Such a situation occurs in the case of the spontaneous symmetry breakdown
(spontaneous magnetization),
where we need a virtual external field to stabilize the system
ensemble.
Then, we should reconsider this case here.

First of all, the occurrence probability of $\Haii$ for the 
temperature $K$ and external field $f$ is given by
\Beq
P(\{\sigma\};K;f) = \frac{\exp(-K H(\{\sigma\})+ f S(\Haii))}{Z(K;f)}.
\label{pdefwithf}
\Eeq
The normalization factor $Z(K;f)$ also depends on the 
external field $f$.
In the mixed-temperature ensemble, the posterior probability $Q(j;\Haii;f)$
of the labeled temperature $K_j$ is 
\Beq
Q(j;\Haii;f)=\frac{P(\Haii;K_j;f)}{\sum_k P(\Haii;K_k;f)}.
\label{qdefwithf}
\Eeq
Therefore, both the maximum accuracy and the guessed 
temperature $j_{\rm max}$ ensuring it depend on 
the external field $f$, and the total theoretical upper limit of the
accuracy depends on $f$.
 
Now, we look at the $\Haii$ dependence of these quantities by
substituting Eq.~(\ref{pdefwithf}) into Eq.~(\ref{qdefwithf}).
The dependence through the total spin function $S(\Haii)$
cancels out between the numerator and the denominator, and
only the dependence through the Hamiltonian survives, 
\Bea
A(\Haii;f) \Rightarrow A(H(\Haii);f),\\
j_{\rm max}(\Haii;f) \Rightarrow j_{\rm max}(H(\Haii);f).
\Eea
We reach the same conclusion again. 
Even in the case where the temperature-independent 
external field also controls the statistical weights, 
the previous conclusion holds, that is, knowing the Hamiltonian
is the {\sl necessary and sufficient} condition for making 
the best available machine.

Note here that what is observed by the
optimized machine does not have to be the Hamiltonian 
itself. Consider a procedure that the machine
observes a physical quantity $X$, obtains a 
value of it, $x$, and guesses the input temperature 
as a function of $x$. If all states having the same value of 
$x$ give a single value of the Hamiltonian, then the
observation of $X$ is equivalent to that of the Hamiltonian.
In other words, this case ensures the existence of
a map $X\rightarrow H$.
Therefore, the observation of such $X$ can also achieve the
highest accuracy.  

\section{Characteristics of Optimized Machine}

In this section, we investigate the optimization condition
of the deep learning machine
to minimize the error function in Eq.~(\ref{errorfunction}).

Let us recall the deep learning machine structure drawn
in Fig.~\ref{machinestructure}.
Our machine, just before the final output stage, 
prepares a single variable $x$ by manipulating the
input spin configuration $\Haii$ through multiple convolutional
layers.

From the discussion in the previous section, the optimized
machine should know the Hamiltonian as a function of
$\Haii$. With this structured machine, the variable $x$
is the unique variable gathering characteristic 
information contained in the input $\Haii$.
Therefore, the variable $x$ must necessarily 
be the Hamiltonian itself 
of the input configuration, $H[\Haii]$, so that the
machine can achieve the highest accuracy and 
the lowest error.
Hereafter, for the perfectly optimized machine,  
we suppose that the variable $x$ 
represents the Hamiltonian $H[\Haii]$ itself.

As mentioned in the previous section, the optimized
machine should know the Hamiltonian or its equivalent $X$,
as long as there is a map from $X$ to $H$, 
to achieve the highest ability.
Therefore, the key variable $x$ does not have to be
the Hamiltonian itself and it can be something else.
However, in later sections, we confirm that, actually, 
the variable $x$ is optimized to represent the 
Hamiltonian itself up to the origin and the unit 
normalization. This is due to the structure of the final 
output layers and the optimization procedure
of our machines, that is, the full connection layer with
linear functions in $x$ is coupled to the softmax
function output and they are optimized to
minimize the error function discussed below, which is
a much stronger condition than 
that bringing the highest accuracy.

Using this variable $x$, the machine makes the 
output $q_j, j=1,2,\cdots 16$, by using the full connection
layer and the softmax function.
First, 16 variables $y_j$ are set by the machine
as defined in Eq.~(\ref{defofyj}),
\Beq
y_j = w_j x + b_j,
\Eeq
where weight $w_j$ and bias $b_j$ are 
the machine parameters to be optimized in this full connection layer.

Then, these variables $y_j$ are transformed
into $q_j$ via the softmax function as in Eq.~(\ref{defofqj}),
\Beq
q_j = \frac{e^{y_j}}{\Sigma_k e^{y_k}}.
\Eeq
These $q_j$ values are the final machine output,
which are interpreted as the machine estimate of the
normalized posterior probabilities of 
input temperature $K_j$  as a function of input $\Haii$.

Before discussing the optimization condition,
we recapitulate the
basic notions of evaluating the difference between
two independent probability functions.
We take
two probability functions $Q(i)$ and $q(i)$, both of
which are normalized,
\Beq
\sum_i Q(i)=1, \sum_i q(i)=1.
\Eeq
We introduce the Kullback -- Leibler (KL)\cite{KL} divergence to represent 
the size of the difference between these two probability functions,
\Beq
D_{\rm KL}(Q||q) \equiv \sum_i Q(i)\log \frac{Q(i)}{q(i)}.
\Eeq
In fact, the KL divergence is straightforwardly 
proved to be non-negative,
\Beq
D_{\rm KL}(Q||q)  \geq 0,
\label{KLnonnegative}
\Eeq
and the equality above is satisfied only at 
$q(i)=Q(i),  {}^\forall i$, which is the unique minimum.

We evaluate the machine by comparing the machine 
estimate of the posterior probabilities with the correct
posterior probabilities predetermined by the input
mixed temperature ensemble.
The predetermined posterior probabilities
$Q(j;\Haii)$ are defined in Eq.~(\ref{posteriorinput}).
Then, the KL divergence comparing this $Q$ and the
machine estimate $q$ for a fixed $\Haii$ is given by
\Beq
D_{\rm KL}(Q||q)[\Haii] 
= \sum_i Q(j;\Haii)\log\frac{Q(j;\Haii)}{q_{j}(\Haii)},
\label{KLdef}
\Eeq
where the machine estimate $q_j$ is calculated for
each input $\Haii$ and we explicitly express the $\Haii$
dependence of $q_j$.

For $Q(j;\Haii)$, although it is predetermined 
by the input mixed-temperature ensemble, the machine
should not be assumed to know it. There is 
a standard practical way of
evaluating the KL divergence in the course of
optimization. We define the error function $Er(j_{\rm in};\Haii)$
for each input $\Haii$ with label $j_{\rm in}$ as
\Beq
Er(j_{\rm in}; \Haii)=-\log q_{j_{\rm in}}(\Haii).
\Eeq
Averaging over the input configurations automatically 
sums up all the possible input temperatures
$j_{\rm in}$ and $\Haii$
with correct probabilities of the input ensemble.

Denoting the average over the input 
mixed temperature ensemble by 
$\langle \cdot \rangle_{j_{\rm in},\Haii}$,
the averaged error function 
\Beq
Er=\langle Er(j_{\rm in}; \Haii)\rangle_{j_{\rm in},\Haii}
\Eeq
is rewritten as follows by separating the summation into 
temperature and $\Haii$,
\Bea
\kern-5mm
Er&\kern-4mm=&\kern-4mm\left\langle\sum_j Q(j;\Haii) Er(j;\Haii)\right\rangle_{\Haii}\nonumber\\
&\kern-4mm=&\kern-4mm-\left\langle\sum_j Q(j;\Haii) \log q_j(\Haii)
\right\rangle_{\Haii}\nonumber\\
&\kern-4mm=&\kern-4mm\left\langle D_{\rm KL}(Q||q)[\Haii] 
-\sum_j Q(j;\Haii)\log Q(j;\Haii)
\right\rangle_{\Haii},
\Eea
where the KL divergence for a fixed $\Haii$ is defined in 
Eq.~(\ref{KLdef}).

The optimization of the machine is performed 
to minimize the error
function by adjusting the output $q_j(\Haii)$ while
$Q(j;\Haii)$ is the fixed input.
Owing to the non-negative condition in Eq.~(\ref{KLnonnegative}),
the minimum error function is realized at
\Beq
q_j(\Haii)=Q(j;\Haii),{}^\forall \Haii, 
\Eeq
and the minimum value is 
\Beq
Er_{\rm min}= -\left\langle \sum_j Q(j;\Haii)\log Q(j;\Haii)\right\rangle_{\Haii}.
\Eeq

Introducing the probability function $R(\Haii)$ for 
$\Haii$  to appear in the total mixed temperature ensemble,
we have
\Bea
Er_{\rm min} &=& 
-\sum_{\Haii} R(\Haii) \sum_j Q(j;\Haii)\log Q(j;\Haii) \nonumber\\
&=&-\sum_{j,\Haii} U(j;\Haii)\log U(j;\Haii) \nonumber\\
&&+\sum_{\Haii} R(\Haii) \log R(\Haii),
\label{minimumerrorfunction}
\Eea
where $U(j;\Haii)$ is defined as
\Beq
U(j;\Haii)\equiv R(\Haii) Q(j;\Haii).
\Eeq
This is the product probability of the temperature $K_j$ and $\Haii$
in the input ensemble.
Therefore, the minimum error function is 
the Shannon entropy of the total input ensemble minus
the remaining entropy ignoring the temperature information.

Here, we should mention the accuracy 
given by the minimum error machine.
The accuracy is frequently used to evaluate the
ability of a machine to discriminate the input objects.
In our case, it is defined by the frequency that the predicted temperature
coincides with the labeled input temperature.
To obtain the highest accuracy, the machine must predict
the temperature $j_{\rm out}$ so that it maximizes
$Q(j;\Haii)$ as in Eq.~(\ref{argmaxaccuracy}),
\Beq
j_{\rm max}(\Haii) = \argmax_j Q(j;\Haii).
\Eeq
This is the best result and no machine can perform better.
As explained earlier, 
the optimized machine should output $q_j(\Haii)$ exactly equal to 
$Q(j;\Haii)$ for all $\Haii$ 
and thus it achieves the highest possible accuracy.

Finally, in this section, we clarify how we should evaluate 
the level of machine power, that is, 
how near the obtained machine is to the 
optimized one. One way is to compare the achieved accuracy 
with the theoretical highest value.
However, the optimization procedure 
is controlled by minimizing the error function.
Therefore, we should evaluate the machine performance
by comparing the error function with the theoretical minimum 
error value obtained in Eq.~(\ref{minimumerrorfunction}).

The value $Er_{\rm min}$ can be evaluated by
MC simulation of the system.
However, if we know the free energy of the
system, it can be calculated simply.
The probability functions 
$R(\Haii)$ and $Q(j;\Haii)$ in Eq.~(\ref{minimumerrorfunction}) 
are expressed as functions of
$\Haii$. However, they are actually functions of 
the Hamiltonian only as stressed in the previous section,
\Beq
R(\Haii) \Rightarrow R(H(\Haii)) ,\ 
Q(j;\Haii) \Rightarrow Q(j:H(\Haii)).
\Eeq
Then, defining the probability density function  $V(E)$ ($E$ is
a value of the Hamiltonian function $H$), we have
\Beq
Er_{\rm min}
=-\int dE V(E) \sum_j Q(j;E)\log Q(j;E).
\label{howtoevaluateerror}
\Eeq
With this expression, we can evaluate $Er_{\rm min}$
straightforwardly if we know the energy expectation 
value and the specific heat at every temperature.
That is, we use the normal distribution approximation
and replace the density function $V(E)$ with 
sum of the Gaussian distributions for each temperature.
Of course, this is not completely correct, but
it turned out to be a practically excellent 
approximation in our statistical models of spins.

\section{Relations between Optimized Machine Parameters
and Physical Quantities}

In this section, we prove the main statement of this article: the 
relations between the optimized machine parameters and the
physical quantities of the statistical system.

As proved in the previous sections, our optimized machine 
must have two features as follows:
\begin{enumerate}
\item The intermediate variable $x$ represents 
the Hamiltonian of the system
 $H(\Haii)$ for all $\Haii$.
\item The output $q_j(\Haii)$ is equal to $Q(j;\Haii)$ for 
all $j$ and $\Haii$.
\end{enumerate}
The purpose of this section is to prove the relations declared in
Eqs.~(\ref{wopt}) and (\ref{bopt}) from these two conditions.

To make the notation simple, we denote $H(\Haii)$ by $E$.
Note again 
that  in the second statement, the condition ``for all $\Haii$" 
is equivalent to ``for all $E$" since $q_j(\Haii)$ and $Q(j;\Haii)$
can be reduced to $q_j(E)$ and $Q(j;E)$, respectively.

First of all, we express the key variable $x$ in terms of $E$,
\Beq
x(E) = -\frac{1}{a_1} E - \frac{c_1}{a_1},
\label{xsetting}
\Eeq
where $a_1$ and $c_1$ are constants. The reason for the somewhat 
peculiar definition here is to simplify the final formula.
We have to allow freedom of the unit normalization
and the origin of energy.
These two constants are irrelevant for physics
and, in fact, the machine can become perfect with these
uncertain arbitrary parameters. This means that the optimized 
machine parameters have completely flat directions for 
the error function.

The full connection layer sets the $y_j$ variable 
defined in Eq.~(\ref{defofyj}). The final output $q_j(E)$
has the relative ratio
\Beq
q_j(E) \propto \exp(w_j x(E)+ b_j),
\Eeq
where the proportionality means 
that as the $j$-space vector (16-dimensional), 
the left-hand-side vector is proportional to the right-hand-side vector.
Next, the predetermined probability $Q(j;E)$
defined in Eq.~(\ref{posteriorinput}) satisfies
\Beq
Q(j;E) \propto P(E;K_j) 
= \frac{1}{J}\frac{\exp(-K_j E)}{Z(K_j)}\ .
\Eeq

Then, the second statement is simply
the following proportionality,
\Beq
\exp(w_j x(E)+ b_j) \propto \frac{\exp(-K_j E)}{Z(K_j)}.
\Eeq
The proportionality factor can depend on $E$ and 
we obtain the equality
\Beq
\exp(w_j x(E)+ b_j) = C(E) \frac{\exp(-K_j E)}{Z(K_j)}.
\Eeq
Taking the logarithm of both sides of this equality,
we have
\Beq
w_j x(E)+ b_j = \log C(E) -K_j E + F_j,
\label{basicequality}
\Eeq
where $F_j$ is the free energy defined by
\Beq
F_j \equiv -\log Z(K_j).
\Eeq

The $E$-dependent proportionality factor $C(E)$
is restricted. Since the above equality holds for
any $E$, $\log C(E)$ must be a linear function in $E$.
We take the following notation:
\Beq
\log C(E) = a_0 E + c_0 - \frac{a_0c_1}{a_1},
\Eeq
where we introduced additional constants
$c_0$ and $c_1$. These constants also represent the
completely flat directions of the error function of the
optimized machine.

Substituting $x$ defined in Eq.~ (\ref{xsetting}) into 
Eq.~(\ref{basicequality}), we have two equalities,
\Bea
w_j &=& a_1 K_j + a_0 ,\label{optweightfunction}\\
b_j &=& F_j + c_1 K_j + c_0.\label{optbiasfunction}
\Eea
Thus, we reach the key result. The full connection layer
parameters of the optimized machine can be given 
by this solution. 
This solution has four free parameters, 
which correspond to the completely flat directions
of the error function.
Of course, these four constants are totally undetermined 
and physically insignificant parameters.

Here, we should mention that the machine does not know
the temperature value $K_j$ itself and just knows the temperature
label $j$. In other words, there is no way or chance to tell
the machine the temperature value $K_j$.
However, the optimized machine learns enough to
know the temperature values through $w_j$ up to
the unit normalization and the origin. These free parameters
do not affect the error function at all.

Let us explain how to calculate the physical singular 
behavior of the specific heat of the statistical system
by using the optimized machine parameters only.
First of all, we reorder the temperature label $j$
so that the weight $w_j$ is monotonic.
The correct direction, 
increasing or decreasing, does not matter in this
stage, which will be clarified later.
Then, we calculate the difference 
\Beq
\Delta_j = w_{j+1} - w_{j},
\Eeq
which will be proportional to the corresponding
spacing of the temperatures,
\Beq
\Delta_j  = a_1 (K_{j+1} - K_j).
\Eeq
Note that we do not assume equal spacing of
the input temperatures.

Then, we evaluate the first derivative of 
the free energy as follows: 
\Beq
\left. \frac{dF}{dK}\right |_{K_j} = -\langle E \rangle_{K_j}
\Longrightarrow \frac{F_{j+1}-F_{j-1}}{K_{j+1}-K_{j-1}}
=a_1 \frac{b_{j+1}-b_{j-1}}{\Delta_{j-1}+\Delta_{j}}-c_1.
\label{biasfirstdifference}
\Eeq
Now, we have the energy expectation value at temperature
$K_j$. The arbitrary constant $c_1$ comes from the fact that
the origin of the energy cannot be determined and
the arbitrary $a_1$ corresponds to the unit 
normalization.  Although even the sign of $a_1$ is indefinite,
we can fix the increasing direction of temperature
so that the energy expectation values 
increase when the temperature increases, 
assuming the normality of the input statistical system.

Finally, we  evaluate the second derivative of the
free energy,
\Beq
\frac{d^2F}{dK^2}=\langle (\Delta E)^2 \rangle_{K_j}
\Longrightarrow 2a_1^2 \left(\frac{b_{j+1}-b_j}{\Delta_j}
+\frac{b_{j-1}-b_j}{\Delta_{j-1}}\right)/(\Delta_{j-1}+\Delta_j).
\label{biasseconddifference}
\Eeq
This quantity must be non-negative and equal to the
specific heat of the statistical system. Therefore, we 
find a peaked structure around the phase transition
point if any.

To summarize this section, we stress the interesting point
of our results. The free energy of the statistical system 
is not a simple quantity to calculate. According to the
definition, we need the partition function, which is
usually impossible to calculate.
However, the deep learning machine, looking at many 
configurations from a mixed-temperature ensemble and
minimizing the error of predicting the input temperature label, 
finally engraves the free energy onto the bias parameters, 
even as a function of temperature.
Investigating these engraves parameters, we can
obtain the specific heat as a function of temperature 
and find the existence of a singularity related to 
the phase transition.

\section{Spin Models to be Learned}

Hereafter, we set up deep learning machines and  make them 
learn the input temperature from input spin configurations.
The spin models we use here are the 
two-dimensional nearest-neighbor Ising  model  on the
square lattice (2d-NNI)
and the one-dimensional long-range Ising model (1d-LRI).
As before, the statistical weight is given by the Hamiltonian $H(\Haii)$,
\Beq
\exp (-K H(\Haii)),
\Eeq
and we call $K$ the temperature (actually the inverse temperature).

The 2d-NNI model is the common nearest-neighbor interaction 
model with the Hamiltonian
\Beq
H = -\sum_{\rm n.n.} \sigma_i\sigma_{i+1}.
\Eeq
This model, in the infinite volume limit, has the second-order phase transition at 
$K=\frac{1}{2}\log (1+\sqrt{2})\simeq 0.44$.\cite{Onsager}
The lattice size is the $32\times 32$ and the
input temperatures are 16 classes in the period
$K=[0.24,\ 0.54]$ with equal spacing of $0.02$.
We expect that the deep learning of 
this finite-volume system should find
the specific heat singularity remnant at the
phase transition temperature.
This model has an exact expression for the free energy 
as a function of $K$. However, it is for an infinite-volume system,
and suffers from a large difference compared with our finite-volume 
system. Therefore, to calculate theoretical 
key quantities to evaluate the 
deep learning machines, we need high-precision MC calculation.

The 1d-LRI model has the long-range interactions
\Beq
H=-\sum_{i,n}K_n \sigma_i \sigma_{i+n},
\Eeq
where the coupling constant $K_n$ is defined by
\Beq
K_n=\frac{1}{n^p}.
\Eeq
The constant $p$ determines the damping rate of the 
interactions. It is known that for $1<p\leq 2$, there is a 
phase transition\cite{D.Ruelle,Griffiths,Griffiths1967_1,dyson1969} at finite $K$.
This model is the most primitive model of quantum 
dissipation, which is an effective theory of the coupled 
harmonic oscillator system after the environmental
oscillators are integrated out. In fact, the long-range
interactions simply mean the nonlocal 
interactions in the time direction.

According to finite range scaling (FRS)\cite{FRS}
analysis, the phase
transition point can be investigated by calculating a
finite-range system where the maximum distance of
interactions is limited to some number.
Although such a finite-range system does not bring about
any phase transition, we can extrapolate the results by
increasing the maximum range to guess the infinite-range
phase transition point.

Here, we set $p=1.8$ and the maximum range to be 8.
We take the lattice size 1024, and the input temperatures
are 16 classes in the period 
$K=[0.2,\ 0.5]$ with equal spacing of $0.02$.
For $p=1.8$, the infinite-range and infinite-volume system has
a phase transition at $K=0.41$.
In this finite-range and finite-volume system, 
a peak of the specific heat, which is a remnant of 
the phase transition, is expected
to appear.

This type of finite-range 1d-LRI model has a good
feature. Formulating the block decimation renormalization 
group (BDRG),\cite{FRS} we can exactly 
calculate the free energy as a function 
of temperature $K$ for any finite-volume system.
Therefore, we can calculate the key 
quantities to evaluate machines almost exactly without 
relying on MC simulation. 
This is a major benefit compared with the 2d-NNI case.

Also, the observation of the phase transition
 must be more difficult in 1d-LRI than in 2d-NNI
since the specific heat singularity of 1d-LRI with this size 
of maximum range is very much smaller than that in the 2d-NNI case.
For these reasons, we take 1d-LRI with a fixed maximum range
as the main model in the following sections 
to investigate the phase transition search 
by the deep learning machine as below.

\section{Deep Learning Machine Details}

In this section, we briefly explain the structure of the deep learning machine.
We take the standard type of convolutional neural 
network\cite{fukushima,Y.LeCun}  with multiple layers.

The input configurations are made by MC simulation. 
For 1d-LRI, we have 
developed a new method of generating configuration data without
relying on the Markov chain procedures. This new method, called 
exact restricted Boltzmann machine (ERBM), generates
configurations with exactly zero autocorrelation.
For 2d-NNI, we adopt the Wolff cluster algorithm\cite{Wollf} to 
make the ensemble stable even below the critical temperature.

In the results reported here, we do not add an external magnetic 
field at all. We have checked that even with an external magnetic
field, all the results we claim still hold as they are.

The number of input configurations is 128,000 for 1d-LRI and 64,000
for 2d-NNI at each temperature. 
For 1d-LRI, we used two types of data, one is the
spin configuration representing spin up or down at each site, and the other
is the domain wall configuration, a dual variable  
representing the existence of domain walls (spin flips). 
The domain wall representation gives generally better results with
quick convergence, and hereafter, we show the domain wall 
representation results.

We work with the Tensorflow software.\cite{google}
In accordance with the standard discipline, we use 80\% of data 
for training and 20\% for testing.  

As explained in the previous sections, we use the error (cost) 
function in Eq.~(\ref{errorfunction}) to drive the stochastic descent 
optimization. Also, the accuracy is observed to check the machine 
quality.

We recapitulate the total machine structure in a very simple notation.
We prepare the input data set  $\mathcal{X}$ and the machine makes
the output data set  $\mathcal{Y}$. 
Our purpose is to design function $M$,
\begin{equation}
M: \mathcal{X} \rightarrow  \mathcal{Y}.
\end{equation}
The machine consists of multiple layers and one layer converts
the input data into the output data as follows:
\begin{equation}
z_i\left({\bf W},{\bf b},{\bf x} \right)=f\left(u_i \right)   ,
\ u_i=\sum_{j} \omega_{ij} x_j +b_i,
\end{equation}
where $f$ is called the activation function, $b$ is bias, 
and $\omega$ is weight.  
The boldface variables {\bf W}, {\bf b}, and {\bf x}
denote matrix or vector of components $b$, $\omega$, and $x$ respectively.

Building up multiple layers, we have
\begin{equation}
z_i^{(l)}\left({\bf W}^{(l)},{\bf b}^{(l)},{\bf x} \right)
=f\left(u_i^{(l)} \right)   \ , \ u_i^{(l)}
=\sum_{j} \omega_{ij}^{(l)} z_j^{(l-1)} +b_i^{(l)},
\end{equation}
where $l$ is the layer number and $z_j^{(0)}=x_j$.
The convolutional layer is characterized through filters as 
\begin{equation}
u_{i,k}=\sum_{p=0}^{H-1}h_{p,k}z_{Si+p,k}+b_{i,k},
\end{equation}
where 
$h$ is a filter of size $H$; the 
subscript $k$ is the channel component and $S$ is the stride. 

By convolution, the input data of size $L$ are
converted into the output data of size 
$(\lfloor \frac{L-H}{S} \rfloor+1)$.
At the final layer, we take the summation (average) 
of all variables in the direction
of the space and channel,
\begin{equation}
x=\sum_i \sum_k z_{i,k},
\end{equation}
where we denote this key variable by $x$, which no longer has a suffix.

Then, we set the full connection layer and apply the softmax function. 
The output of the softmax function is interpreted as
the probability that the input configuration belongs to the $j$th class, 
\begin{equation}\begin{aligned}
P(C_k|{\bf W},{\bf b},{\bf x})&=\softmax_k(u_1,u_2,\ldots,u_K)\\
&\equiv \frac{e^{u_k}}{\sum_l e^{u_l}}  \ ,\  u_k=\omega_{k} x +b_k .
\end{aligned}\end{equation}
The output $y$ is class $k_{\rm max}$ giving the maximum probability,
\Beq
{y}\left({\bf W},{\bf b},{\bf x}\right)=\argmax_k P(C_k|{\bf W},{\bf b},{\bf x}).
\Eeq

To optimize the machine, we minimize 
the error function equivalent to the cross entropy,
\begin{equation}
E({\bf W},{\bf b}) =-\sum_{n=1}^N\sum_{k=1}^K\left\{\delta_{y_nk} \log\left(  {y}\left({\bf W},{\bf b},{\bf x}_n\right) \right)  \right\},
\end{equation}
where $n$ represents the configuration number and $N$ is the
total number.
The machine performance is also evaluated through the accuracy $A$
defined by
\begin{equation}
A=\frac{1}{N}\sum_{n=1}^N\delta\left(  { y}\left({\bf W},{\bf b},{\bf x}_n\right),y_n\right).
\end{equation}

We show our typical setting of the machine parameters (hyperparameters) 
in Table \ref{hyperparameter}.
These hyperparameters  are determined on the basis of experience (trial and error actually).
The parameter $N_{\rm a}$ will be explained in Sec.~7.

\begin{table}
\caption{Hyperparameters (1d-LRI).}
\label{hyperparameter}
\begin{center}
\begin{tabular}{ll}
\hline
\multicolumn{1}{c}{hyperparameter} & \multicolumn{1}{c}{value} \\
\hline
Convolution layer $l$ & $4$ layers \\
layer 1 filter size $H_1$ & $8$  \\
layer 1 stride $S_1$& $8$  \\
other layer filter size $H$ & $2$  \\
other layer stride $S$& $2$  \\
Mini batch size & $120$ \\
Learning rate & $0.0001$ \\
Channel & $8$ \\
$N{\rm a}$ & $1$ , $2$ , $4$ \\
\hline
\end{tabular}
\end{center}
\end{table}

For the activation function, 
we use the rectified linear function (ReLU) , 
\begin{displaymath}
f(x) =\max\{0,x\} =\left\{ \begin{array}{l}
\displaystyle 
x \;\;\;\; ( x \ge 0 ), \\
\displaystyle 
0 \;\;\;\; ( x< 0 ) .
\end{array} \right.
\end{displaymath}
We adopt the AdamOptimizer\cite{Adam} in Tensorflow 
for the stochastic descent method.

\section{Theoretical Evaluation of Maximum Achievement}

\begin{figure}[!h]
\begin{center}
\includegraphics[clip,scale=0.6]{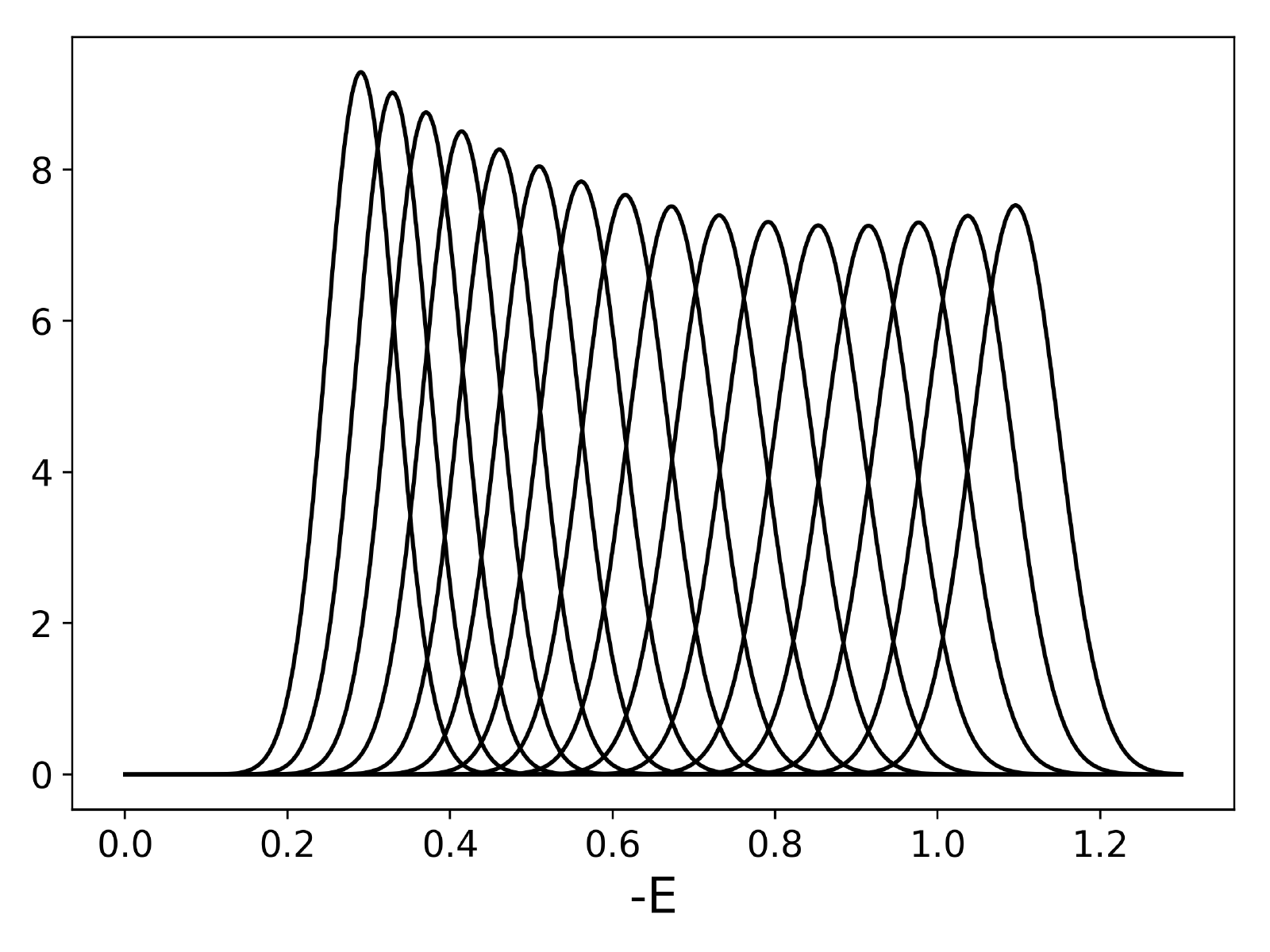}
\caption{Energy distribution (1d-LRI )
from $K=0.2$ (left) to $K=0.5$ (right) in order.}
\label{GaussianLRI}
\end{center}
\end{figure}

\begin{figure}[!h]
\begin{center}
\includegraphics[clip,scale=0.6]{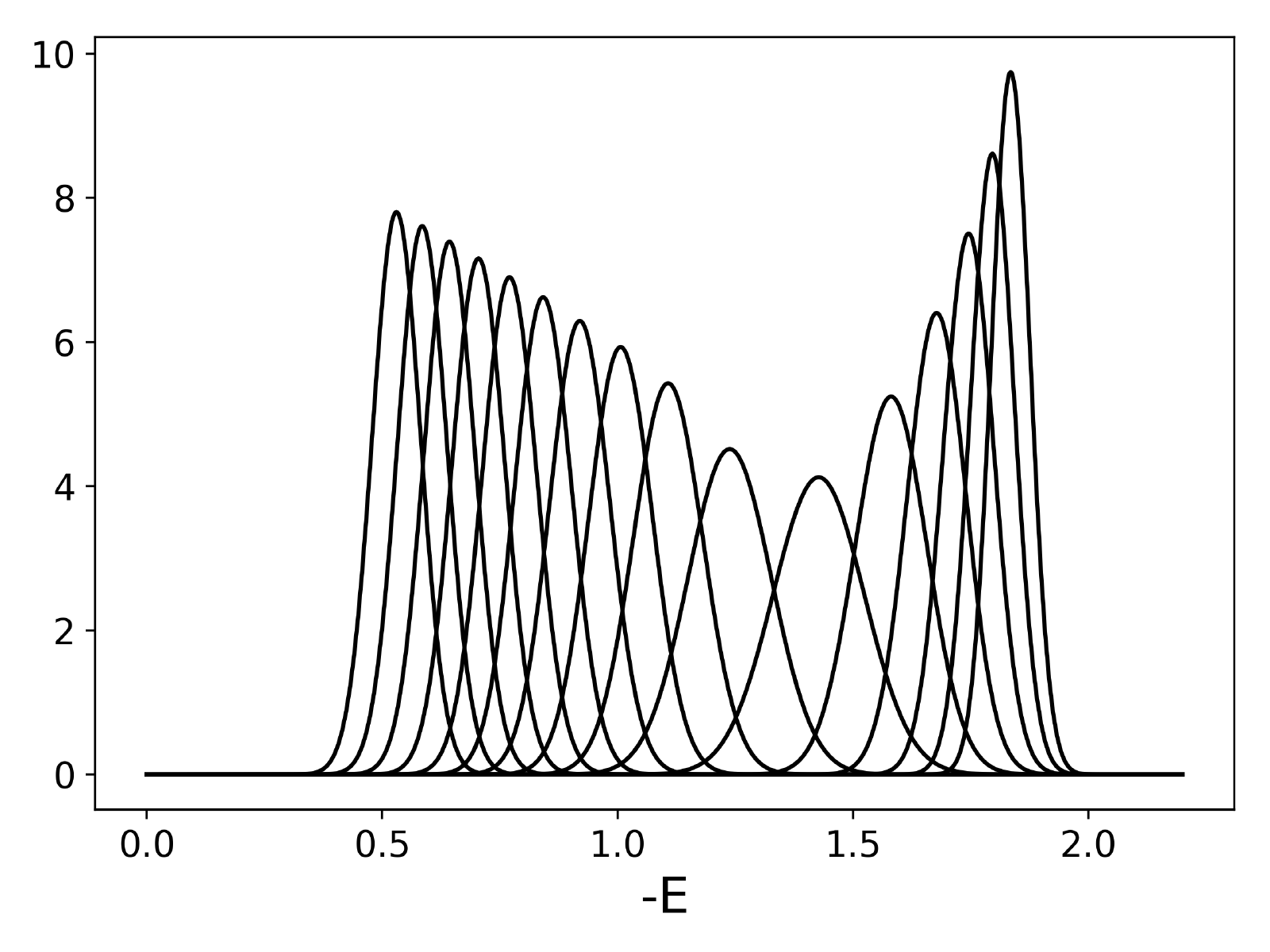}
\caption{Energy distribution (2d-NNI)
from $K=0.24$ (left) to $K=0.54$ (right) in order.}
\label{2dising_gauss}
\end{center}
\end{figure}

Before starting the machine learning, we calculate the 
theoretical limit of the accuracy and the error function so that
we can be sure about the level of optimization.

The possible highest accuracy is given in 
Eq.~(\ref{defoftotalaccuracy}). To understand the meaning
intuitively, we start with a plot of the energy distribution 
for each temperature in the total ensemble. 

In Fig.~\ref{GaussianLRI}, we plot the energy
distribution of 16 temperatures in 1d-LRI. As noted in the
previous section, we can calculate the free energy by
BDRG as a function of temperature for this finite-volume system. 
Therefore, we obtain the energy expectation value and the energy
fluctuation (specific heat) for each temperature. Using these two 
data, we approximate the energy distribution by the Gaussian
form. This figure is plotted using this Gaussian approximation, 
which will be understood as sufficiently precise for our purpose. 

For the 2d-NNI case, we have to perform the MC simulation to calculate 
the energy expectation and fluctuation of a $32\times 32$ finite-volume system for each temperature. Then, using the Gaussian 
approximation, we plot the energy distribution in
Fig.~\ref{2dising_gauss}.

\begin{figure}[th]
\begin{center}
\includegraphics[clip,scale=0.14]{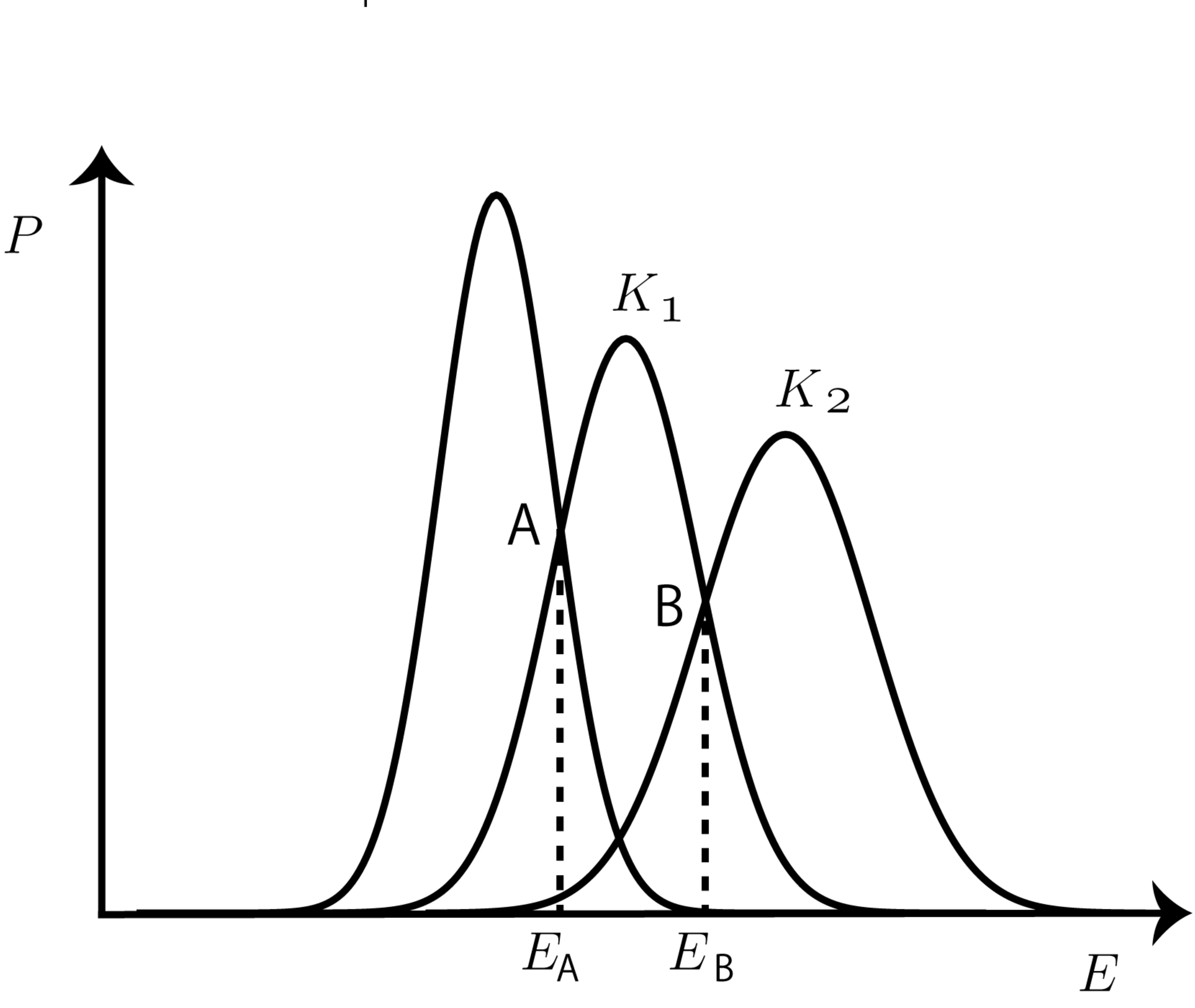}
\caption{Maximum likelihood estimate of the temperature.}
\label{ML}
\end{center}
\end{figure}

We note here that in
Figs.~\ref{GaussianLRI} and \ref{2dising_gauss},
peak structures of the specific heat are already seen.
Since each energy distribution is normalized, a lower
peak position corresponds to a wider distribution, that is, 
a larger specific heat. By comparing these two figures, the 2d-NNI
singularity remnant must be better recognized 
than that of 1d-LRI, by a human at least.

As stressed in Sec.~2, to achieve the highest accuracy, 
the machine should know the energy of the input configuration and then
select the temperature having the largest posterior probability.
In energy distribution plots, this procedure is performed as 
follows: calculate the energy of the input configuration, look 
at the energy distribution plot, 
and select the largest probability temperature.

In Fig.~\ref{ML}, we draw a schematic diagram for the logic of 
the maximum likelihood estimate. For the
configurations whose energy is in the period $[E_{\rm A},E_{\rm B}]$, 
an optimized machine 
should output temperature $K_1$ since it is the maximum
likelihood estimate. Among these configurations, 
those that are judged to output the correct answer 
are only those belonging to the $K_1$ distribution.
Therefore, the total accuracy is simply the area below the upper
envelope of these energy distribution lines. The overlapping 
region contributes to reducing the accuracy.

Thus, the best prediction changes 
at the crossing points of the largest distribution curves.
This transition point is directly calculable.
In Fig.~\ref{ML}, at $E=E_{\rm B}$,
the two neighboring probability functions
 ($K_1,K_2$) coincide at point B,
\Beq
\frac{\exp(-K_1 E_{\rm B})}{Z(K_1)}=\frac{\exp(-K_2 E_{\rm B})}{Z(K_2)}.
\Eeq
Note that the number of configurations with the same energy
does not depend on $K$.
Taking the logarithm of both sides, we have
\Beq
E_{\rm B} = -\frac{F(K_2) - F(K_1)}{K_2-K_1}.
\Eeq
This crossing point can be graphically understood from the plot
of the free energy as a function of $K$.
Recalling that the energy expectation is given by
\Beq
<E>_K = -\frac{dF}{dK},
\Eeq
the crossing point  $E_{\rm B}$ is simply the contact point 
of the tangent with the slope of the line connecting the two points of 
the graph at $K_1$ and $K_2$.
According to the convexity of the free energy as a function of
temperature, we have the inequality (we suppose $K_1<K_2$)
\Beq
\langle E \rangle_{K_1} >E_{\rm B} >\langle E \rangle_{K_2} .
\Eeq

Using the Gaussian approximation,
we can evaluate the area below the upper envelope and
find the theoretical maximum accuracy: 0.43634 (1d-LRI)
and 0.51177 (2d-NNI).  For 1d-LRI, 
we check this accuracy by alternative evaluation
through the very high precision MC simulation, ERBM, and the result is
0.4364(1). This demonstrates that the Gaussian approximation 
is sufficiently precise for evaluating the accuracy limit.

Note that humans, and also machines, probably tend to
look at the spin itself, and may count the total spin
to guess the temperature. If we use the total spin to
discriminate the temperature, then the maximum accuracies 
are only $0.094$ (1d-LRI) and $0.296$ (2d-NNI), which 
are much less than the above upper limit.
Of course, as proved in Sec.~2, any method relying on 
physical quantities other than energy must decrease total accuracy.

\begin{figure}[!h]
\begin{center}
\includegraphics[clip,scale=0.6]{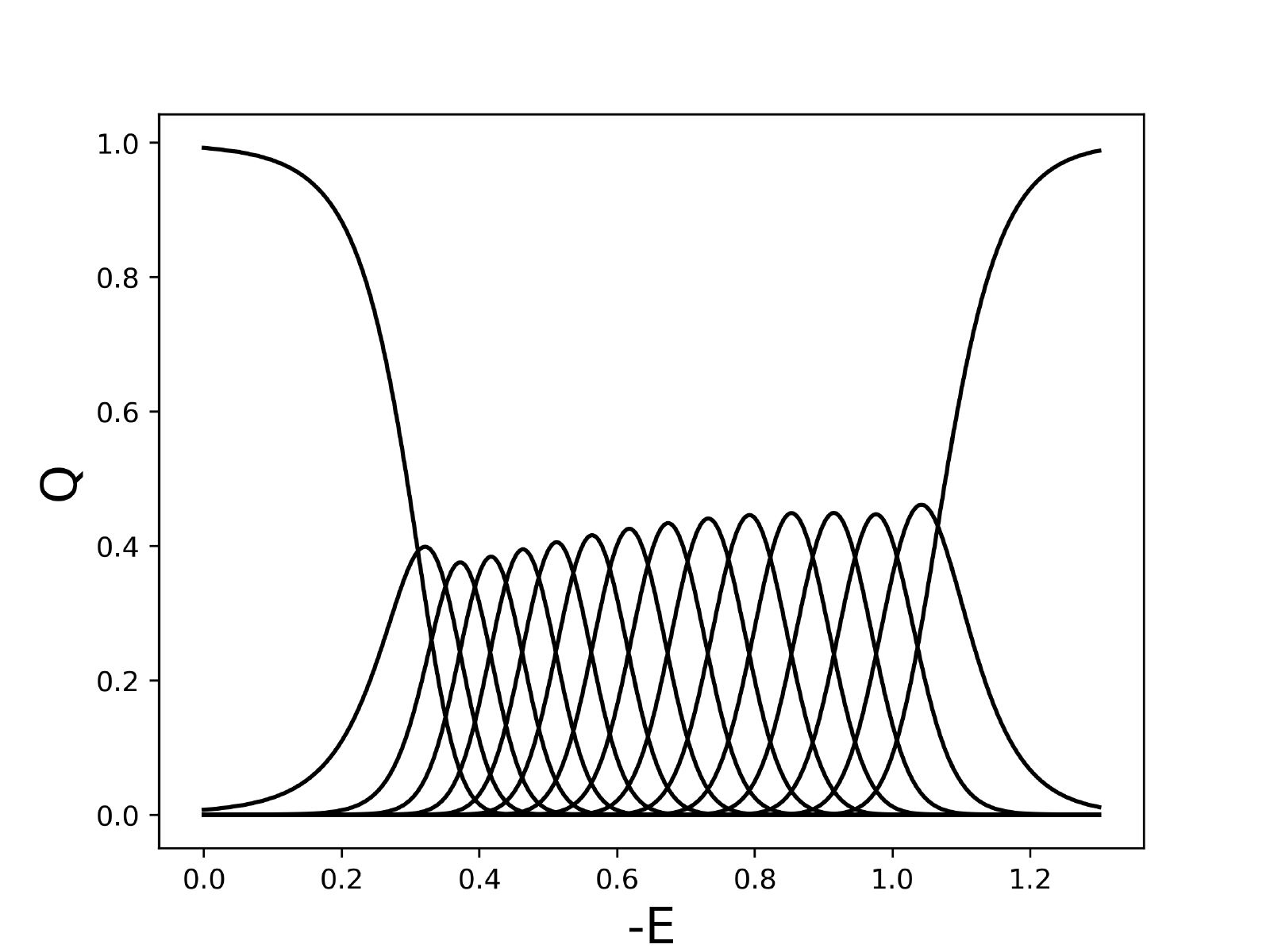}
\caption{Posterior probabilities (1d-LRI)
from $K=0.2$ (left) to $K=0.5$ (right) in order.}
\label{entropydensity}
\end{center}
\end{figure}

Next, we consider the minimum of the error function. 
As written in Eq.~(\ref{howtoevaluateerror}), we can calculate 
it directly from the energy distribution functions.
In Fig.~\ref{entropydensity}, the posterior probability functions
$Q(j;E)$ are plotted for 
each temperature number $j$. We integrate the entropy
$\sum_j Q(j;E) \log Q(j;E)$ with the total energy 
probability density $R(E) dE$ to obtain the minimum error.
The Gaussian approximation gives $1.2697$ (1d-LRI) and $1.0800$ (2d-NNI).
The ERBM high-precision MC simulation gives $1.2704(1)$ for 1d-LRI.
Thus, also for the error function, 
the Gaussian approximation is extremely good.

The theoretical highest accuracies do not appear high.
In order to make them higher, we may bundle input configurations, 
that is, the machine looks at $N_{\rm a}$ configurations at once. 
We denote each 
energy of these $N_{\rm a}$ configurations by 
$E^{(i)}_k$, $k=1,2,\cdots N_{\rm a}$, and $i$ represents
the configuration set number.

Then the probability of occurrence of this bundled configuration
for temperature $K$ is given by the product of the probabilities of 
the component configurations,
\begin{equation}\begin{aligned}
\mathcal{P}\left(\{\sigma\}_i,K\right)&=\prod^{N_{\rm a}}_{k=1}\frac{\exp\left\{-KE_k^{(i)}\right\}}{Z(K)}=\frac{ \exp\left\{-K\sum_{k=1}^{N_{\rm a}}E_k^{(i)} \right\}}{\left(Z(K)\right)^{N_{\rm a}}}\\
&=\left(\frac{\exp\left\{-KE_{\rm a}^{(i)}\right\} }{Z(K)}\right)^{N_{\rm a}}=\left(P(E_{\rm a}^{(i)};K) \right)^{N_{\rm a}}.
\label{bundledconfiguration}
\end{aligned}\end{equation}
Here, we introduced the average energy $E_{\rm a}^{(i)}$ for $N_{\rm a}$ configurations,
\begin{equation}
E_{\rm a}^{(i)}=\frac{1}{N_{\rm a}}\sum_{k=1}^{N_{\rm a}}E^{(i)}_k.
\end{equation}

We see that the average energy $E_{\rm a}$ now controls the
probability. Then the previous calculations hold by replacing 
the energy of the configuration with the average energy of the bundled
configurations.  The average energy distribution functions
have the same expectation value and $1/N_{\rm a}$ of the variance of 
the original single energy distribution. Note that
the crossing point energy between distribution curves does not change
by bundling.
By increasing $N_{\rm a}$, the distribution becomes
narrower and the accuracy must increase. For example, for 1d-LRI, 
we have the upper limits for the accuracy
$0.56852$ ($N_{\rm a}=2$)  and $0.72154$ ($N_{\rm a}=4$), and 
the lowest errors $0.959$ ($N_{\rm a}=2$)  and $0.640$ ($N_{\rm a}=4$).
Thus, the upper bound of the accuracy increases and the lowest error
decreases markedly. This means that by using bundled multiple configurations, 
the amount of information becomes larger and 
the temperature can be discriminated more easily.
However, these machines with $N_{\rm a}>1$ turned out to be worse
for the phase transition search, as  seen later in Sec.~8.

\begin{figure}[!t]
\begin{center}
\includegraphics[clip,scale=0.6]{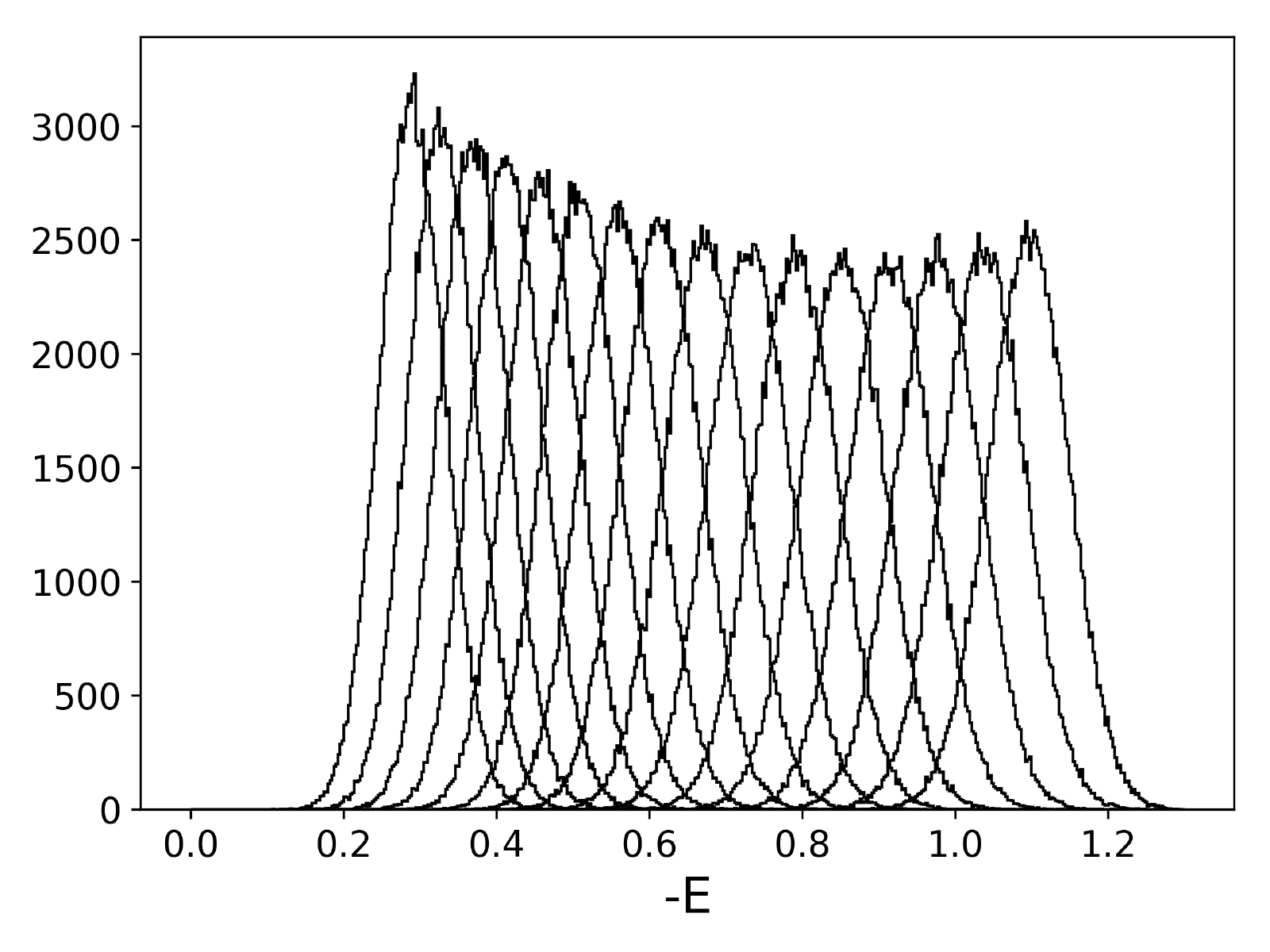}
\caption{Energy distribution of input configurations
from $K=0.2$ (left) to $K=0.5$ (right) in order.}
\label{inputenergydist}
\end{center}
\end{figure}

\begin{figure}[!t]
\begin{center}
\includegraphics[clip,scale=0.6]{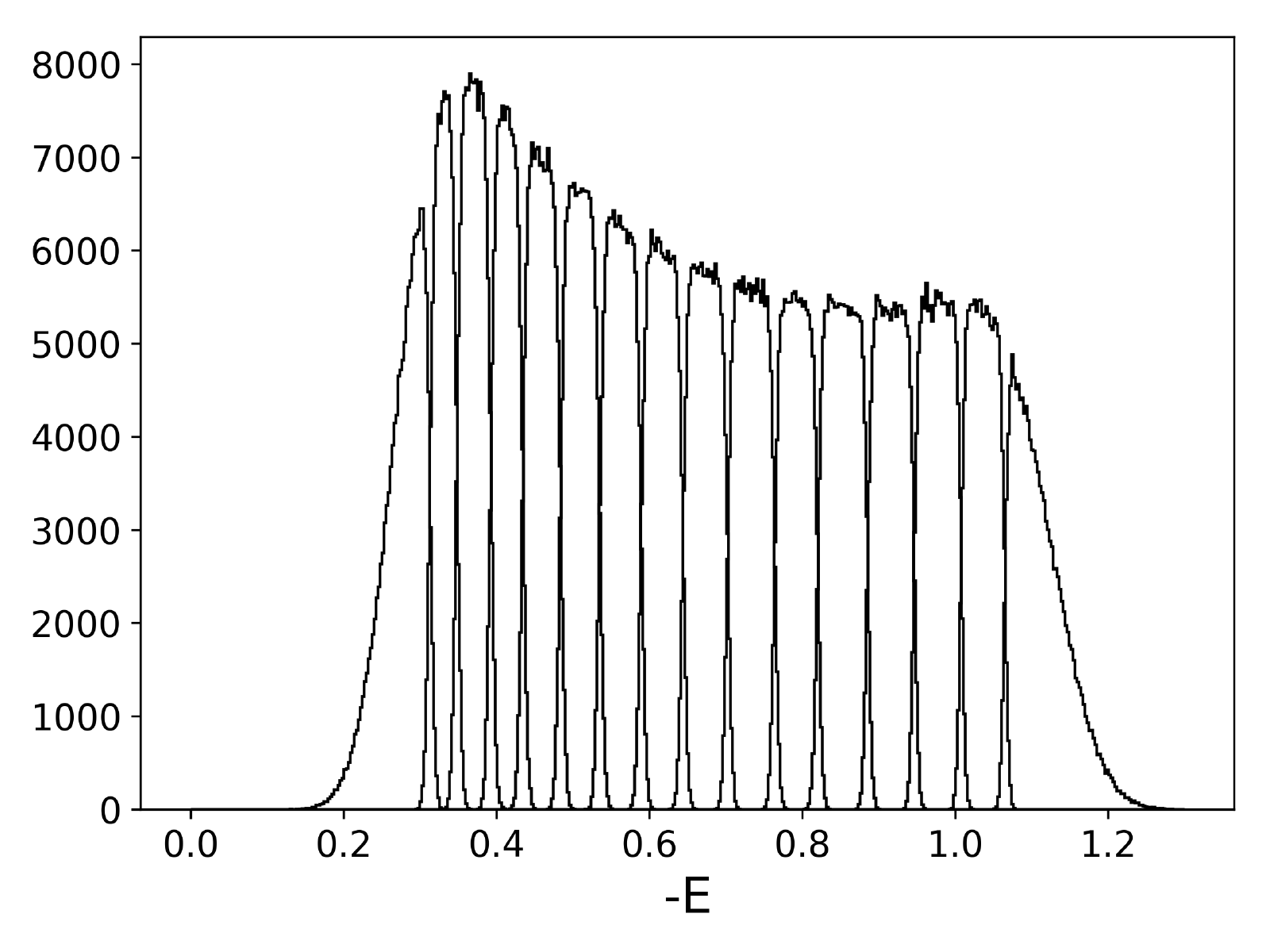}
\caption{Energy distribution for each output temperature
from $K=0.2$ (left) to $K=0.5$ (right) in order.}
\label{outputenergydist}
\end{center}
\end{figure}

\section{Optimizing Deep Learning Machine}

We analyze the optimized machine parameters.
Here, we will report detailed results for the 1d-LRI case.
For the 2d-NNI case, the main results are already shown in 
Sec.~1.

The energy distributions of the input MC configurations are plotted in
Fig.~\ref{inputenergydist}. They reproduce well the Gaussian approximated
theoretical distribution in Fig.~\ref{GaussianLRI}.

For evidence of the level of optimization, we refer to the accuracy and 
the error function achieved. 
For 1d-LRI,  after 2M iterations of learning procedures,
the error function reaches below $1.2755$ and becomes stable, which should be
compared with the theoretical minimum value of $1.2704$. Also, the accuracy 
becomes above $0.435$, whose theoretical upper bound is $0.4364$. 

For 2d-NNI, after 6M iterations of learning procedures, 
the error function is pushed down to $1.111$ where the 
theoretical minimum is $1.0800$,
and the accuracy becomes $0.501$, where the theoretical upper limit is $0.511$.
From these levels of achievement, we understand that
the machines are optimized very well.

The resultant output temperatures are plotted in Fig.~\ref{outputenergydist}.
This figure shows the energy distribution of the configurations, which are 
classified to each predicted temperature.
The shapes of the distributions are far from Gaussian.
The perfect output must show up as rectangular distributions, where
the predicted temperature changes suddenly at the crossing points of
energy. Since the discrimination is not perfect, there are some
overlaps of distribution lines. 

\begin{figure}[!h]
\begin{center}
\includegraphics[clip,scale=0.6]{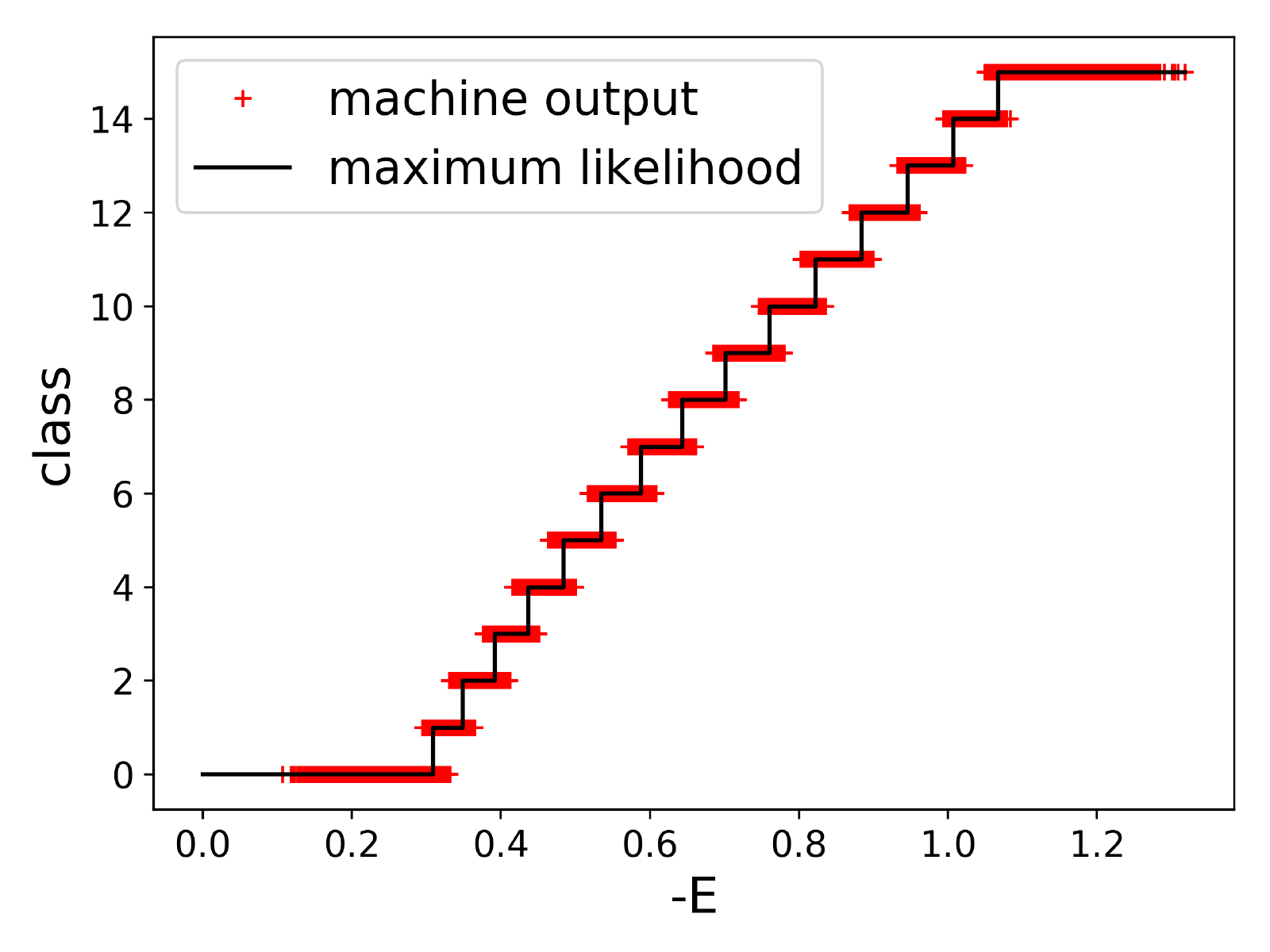}
\caption{Output temperature class vs input configuration energy.}
\label{ene-cla1}
\end{center}
\end{figure}

As stressed in the previous sections, to achieve high-accuracy results, 
we need precise energy observation of input configurations.
Figure \ref{outputenergydist} shows that the energy discrimination is successful
in the sense of energy blocks.
This feature is also verified in Fig.~\ref{ene-cla1}, where we plot the 
relation between the output class and the energy of the configuration. 
The perfect machine gives a contiguous step function 
determined by the maximum likelihood estimate, which is drawn by
a polygonal line there. 

\begin{figure}[!h]
\begin{center}
\includegraphics[clip,scale=0.6]{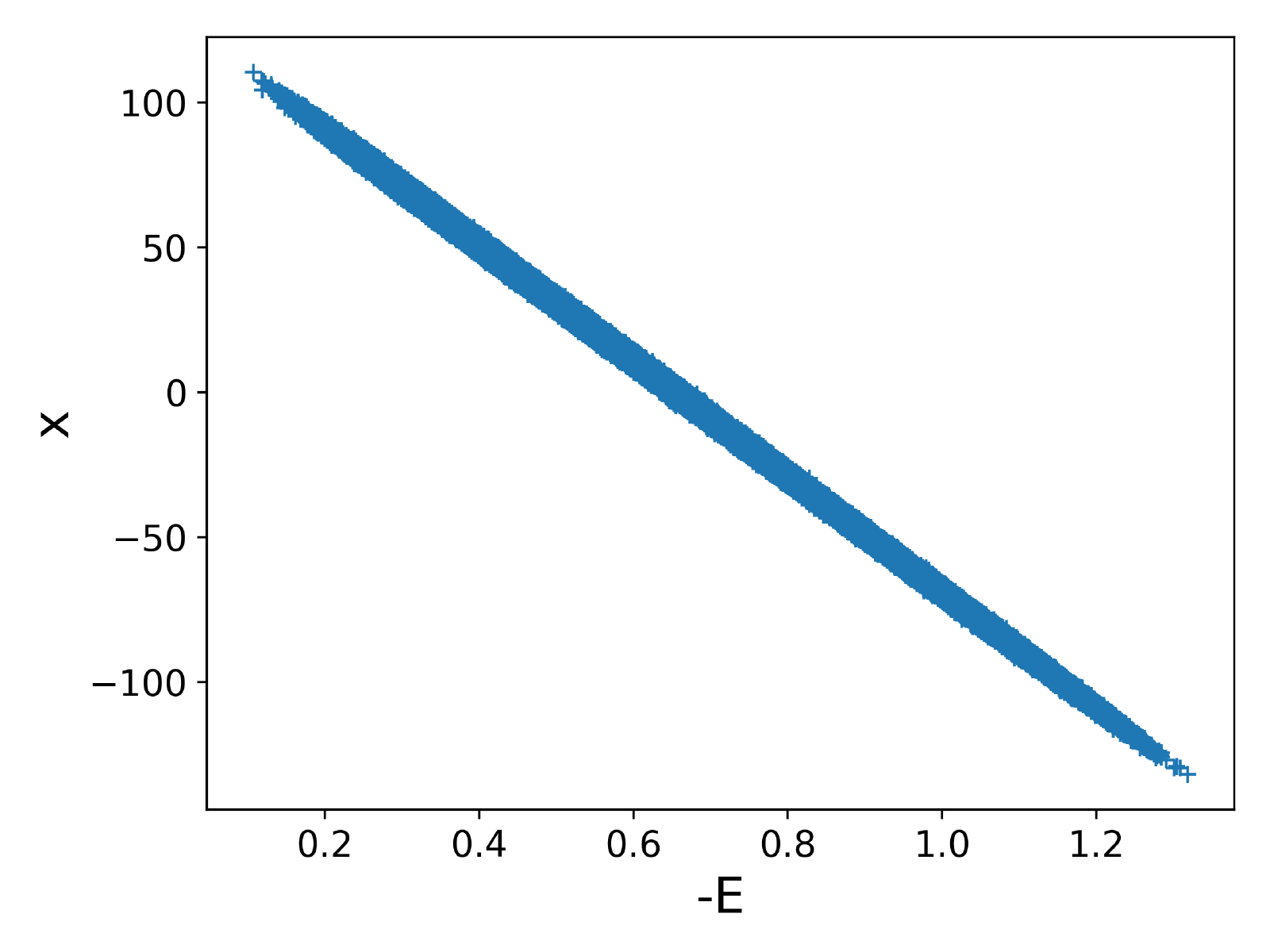}
\caption{Energy vs key variable $x$.}
\label{energyvsx}
\end{center}
\end{figure}

To look into the details of energy discrimination, we examine the 
intermediate key variable $x$. We plot the value $x$ vs the input
configuration energy in Fig.~\ref{energyvsx}. 
This diagram shows the remarkable result that the variable $x$ is 
in fact organized as a linear function of the input energy. 
There is some broadening of the map, which causes the 
remaining errors.

The accuracy itself is not so high, only $0.435$. However, this fact itself
does not matter at all. The central issue is how near this value
is to the theoretical upper limit, which can be realized only by the
exact observation of the energy of the input configuration.
In fact, Fig.~\ref{energyvsx} proves that the machine has learned 
how to evaluate the energy of input configurations. This is 
equivalent to the fact that the machine now knows the Hamiltonian 
function $H(\Haii)$ of the statistical system.

\begin{figure}[!h]
\begin{center}
\includegraphics[clip,scale=0.6]{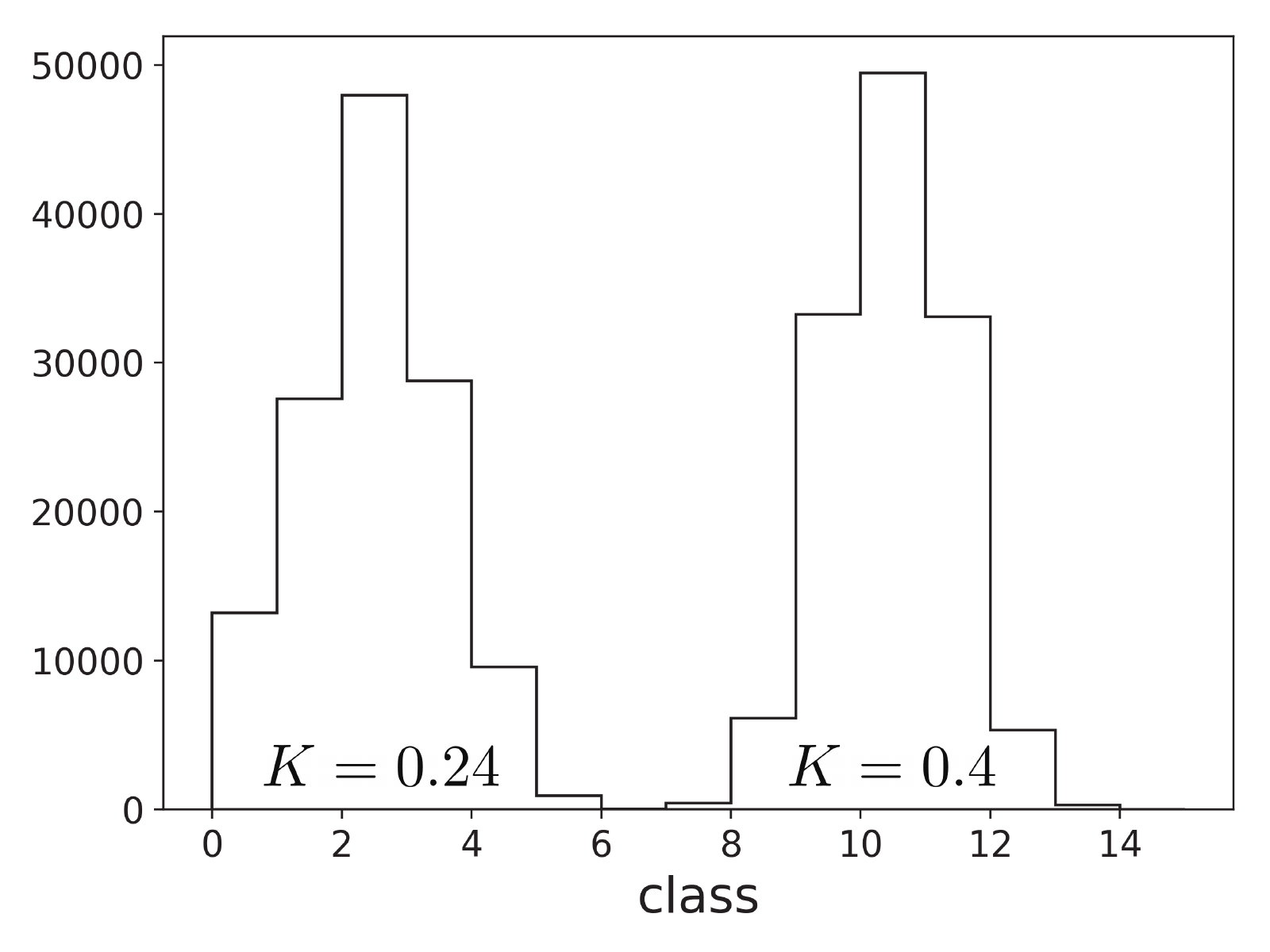}
\caption{Output class histogram for input configurations of a definite temperature.}
\label{spectrometer}
\end{center}
\end{figure}

To make the above point clear, we plot the output histogram 
of the configurations originating from a single input temperature.
Figure \ref{spectrometer} shows the output for the input temperature
$K=0.24$ and $0.4$ configurations.

Let us consider the implications of these histograms.
The central peak of each histogram corresponds to the case that
the output temperature coincides with the input temperature.
The height is not so large and actually around 40\% of the
total histogram, which 
is the accuracy itself. Thus, this diagram might be interpreted to mean that
the machine made mistakes for many configurations, that is,  60\% of
the input configurations are assigned an incorrectly predicted temperature.

Here comes the most important point of this article.
We will answer  the basic
question about the optimized machine, 
that is, what is learned by the deep learning?
The above statement, although it is correct, must be reinterpreted from the 
opposite viewpoint.
That is, this diagram should be understood as {\sl the energy spectrometer}
outputs of the input configurations of a definite temperature.  
Thus, the machine becomes the energy spectrometer after learning.
By supervised learning through the temperature estimate, 
the machine possessed an ability of the spectrometer, the analyzer of the energy.
Note that the energy is chosen automatically since it is the variable conjugate to the temperature.

Now, we claim again that the low accuracy itself does not matter.
The machine has learned much deeper information, the Hamiltonian
of the statistical system.
Now, it is time for the machine to show its obtained power.

\begin{figure}[!h]
\begin{center}
\includegraphics[clip,scale=0.6]{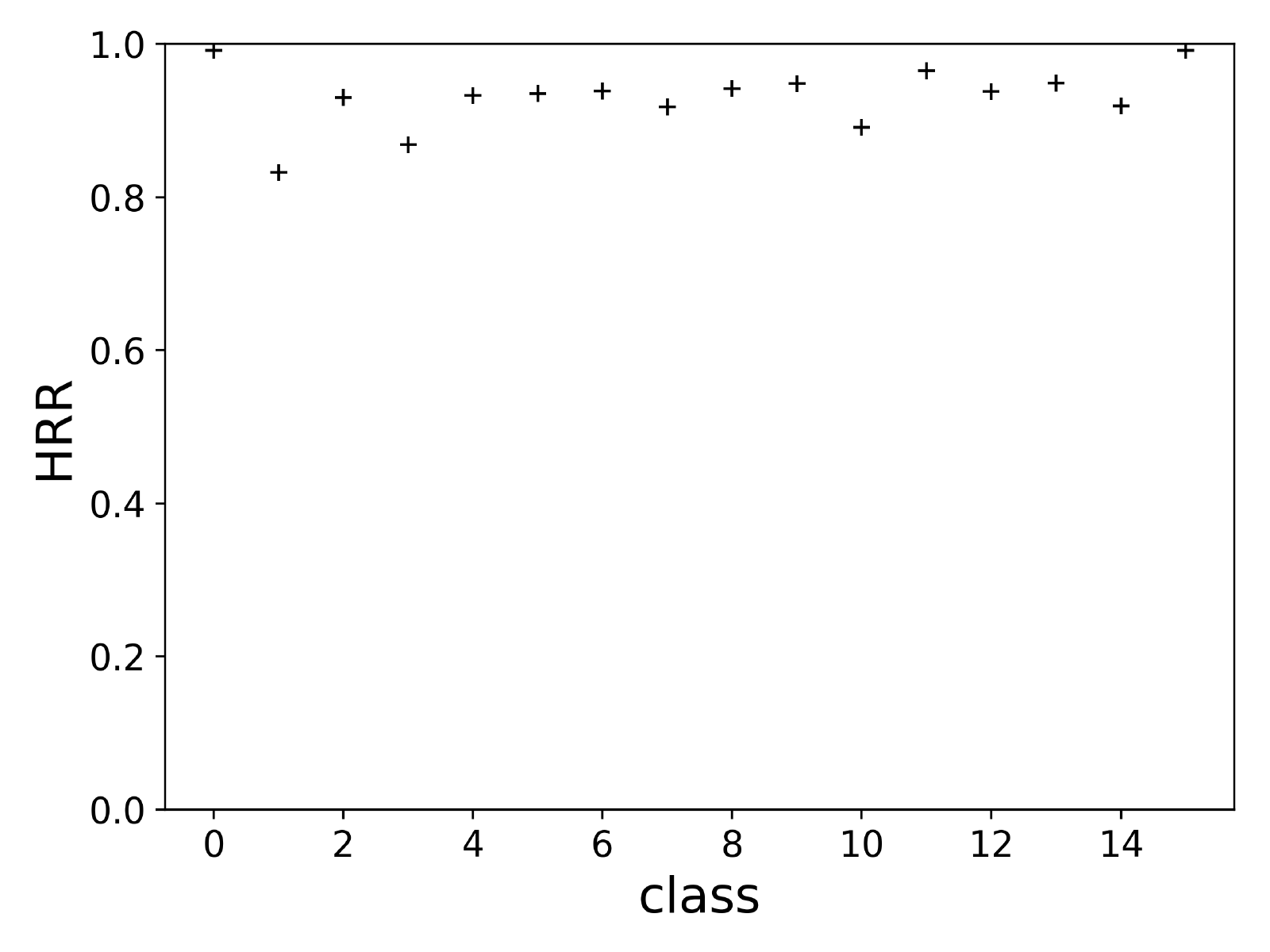}
\caption{Hamiltonian recognition rate: $R=0.937$ (1d-LRI).}
\label{HRR1}
\end{center}
\end{figure}

We define another characteristic to evaluate machine ability, 
the Hamiltonian recognition rate (HRR).
Referring to Fig.~\ref{ene-cla1}, we define the HRR $R_n$ 
for output class $n$ by
\Beq
R_n \equiv \frac{M_n}{N_n},
\Eeq
where $N_n$ is the number of inputs whose best output temperature class is
$n$, and $M_n$ is the number of inputs that are on the correct step 
function line. In Fig.~\ref{HRR1}, we plot the HRR achieved by this 
machine. The total HRR,
\begin{equation}
R=\frac{\sum_n M_n}{\sum_n N_n},
\end{equation}
is $0.937$ for 1d-LRI and $0.867$ for 2d-NNI. 
By evaluating the machine from this viewpoint
instead of the standard accuracy value itself, we can give a better
rating of the optimized machine.
 
Now, we examine the optimized parameters in the full connection
layer. We proved in Sec.~4 that the parameters $w_j$ and $b_j$ 
are expected to be
optimized with arbitrary linear functions of $K_j$ as 
Eqs.~(\ref{optweightfunction}) and (\ref{optbiasfunction}).

\begin{figure}[!h]
\begin{center}
\includegraphics[clip,scale=0.6]{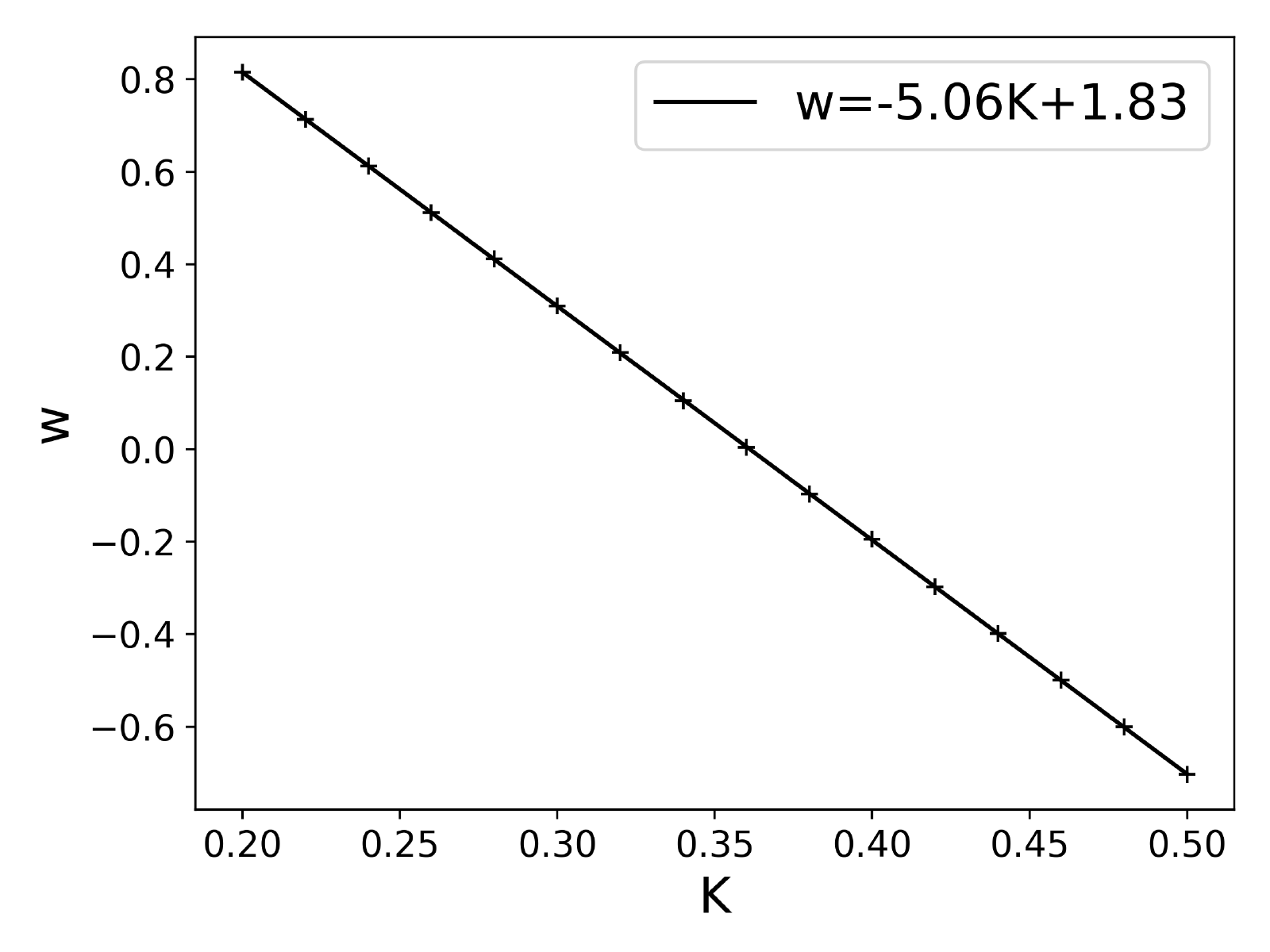}
\caption{Linear dependence of optimized weight $w_j$ on $K_j$  (1d-LRI).}
\label{weightK}
\end{center}
\end{figure}

\begin{figure}[!h]
\begin{center}
\includegraphics[clip,scale=0.6]{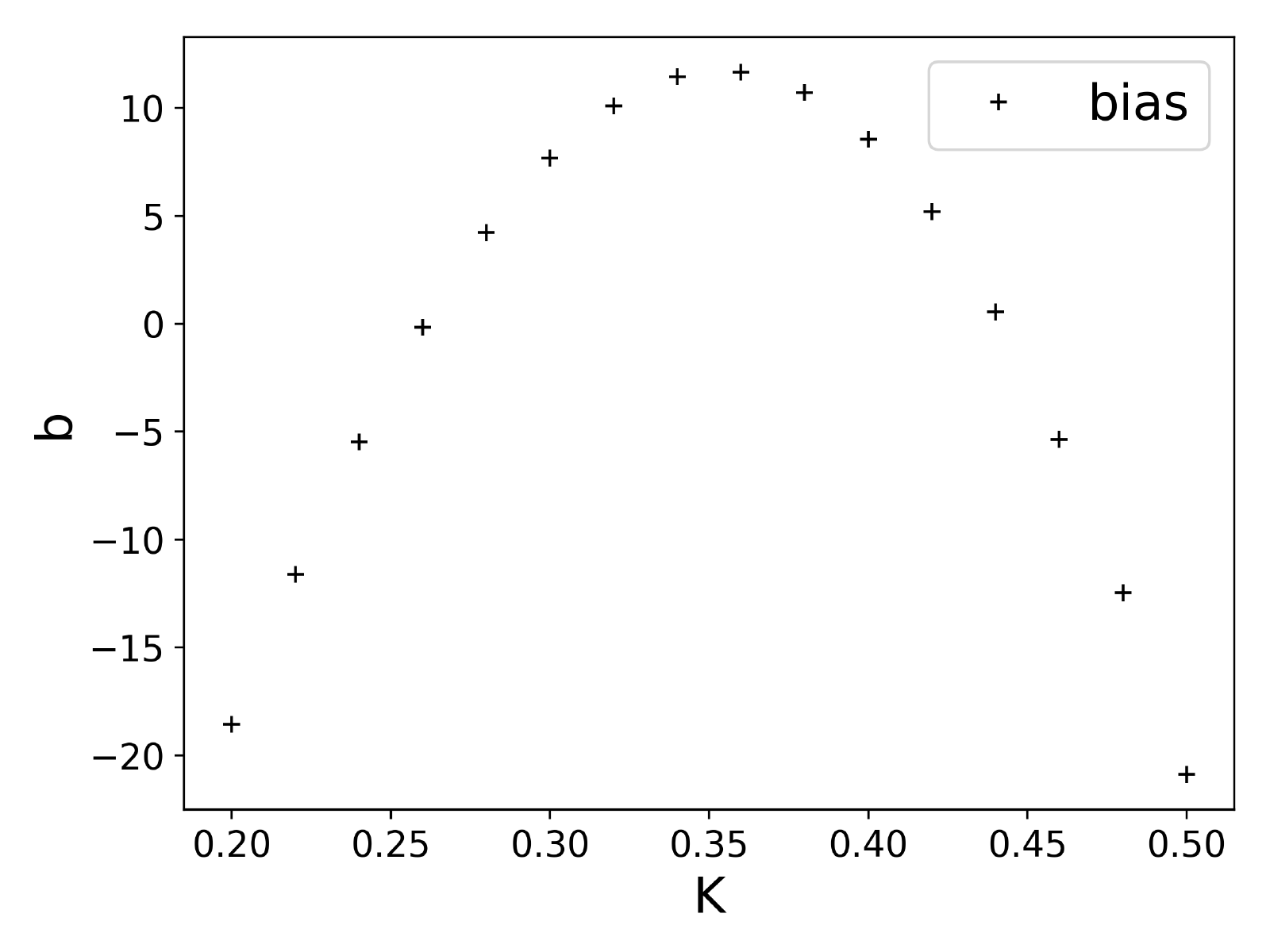}
\caption{Optimized bias $b_j$ vs $K_j$ (1d-LRI).}
\label{biasK}
\end{center}
\end{figure}

\begin{figure}[!h]
\begin{center}
\includegraphics[clip,scale=0.6]{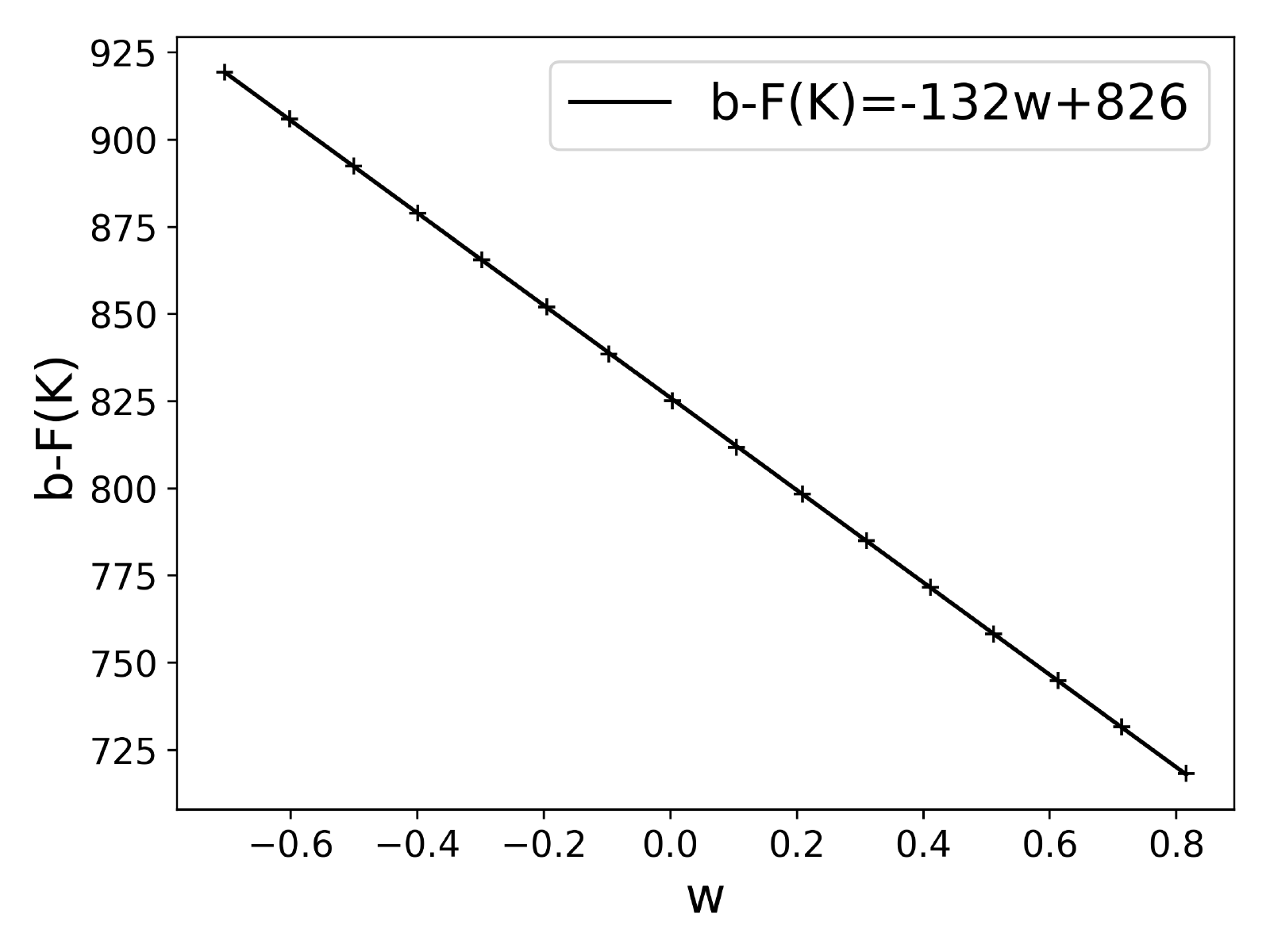}
\caption{Linear dependence of $b_j - F_j$ on $w_j$ (1d-LRI).}
\label{biasFweight}
\end{center}
\end{figure}

In Fig.~\ref{weightK}, the optimized weight $w_j$ in the full connection
layer is plotted with respect to $K_j$. Here, for simplicity, we
take equally spaced temperatures $K_j$ and the diagram axis uses $K_j$ 
instead of their label $j$.
The optimized weight $w_j$ is a linear function of $K_j$, just as we
proved in Eq.~(\ref{optweightfunction}) for the optimized machine.

Next, we plot the optimized bias $b_j$ in Fig.~\ref{biasK}.
It must have a nonvanishing second derivative. 
According to our relation Eq.~(\ref{optbiasfunction}) proved in Sec.~4,
we plot $b_j - F_j$ (bias minus free energy) 
as a function of $w_j$ in Fig.~\ref{biasFweight}.
It shows a perfectly linear function.
Therefore, the optimized machine parameters satisfy the
predicted formula.

\begin{figure}[!h]
\begin{center}
\includegraphics[clip,scale=0.6]{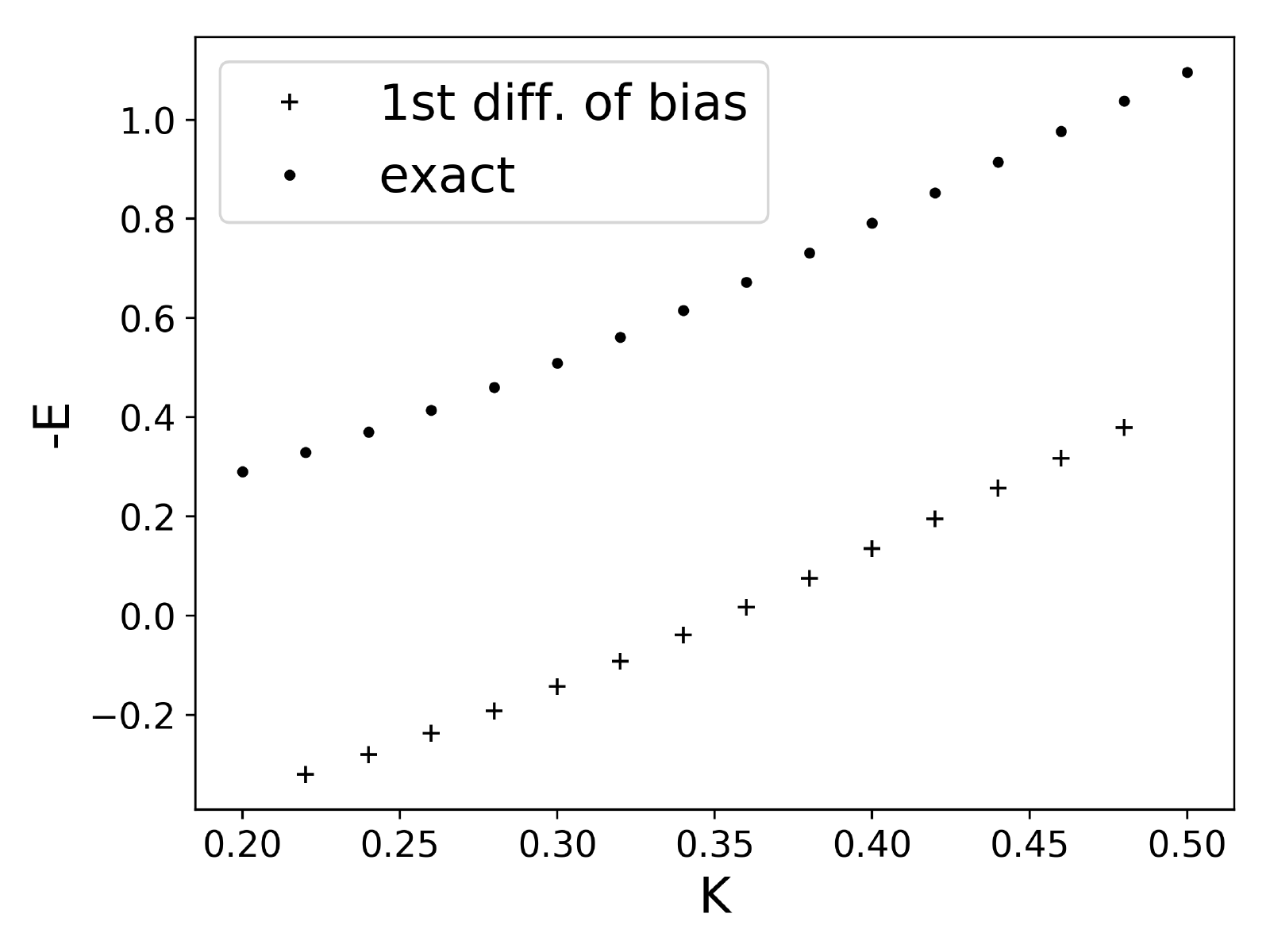}
\caption{First difference of $b_j$ and exact energy expectation value.}
\label{bias1K}
\end{center}
\end{figure}

\begin{figure}[!h]
\begin{center}
\includegraphics[clip,scale=0.6]{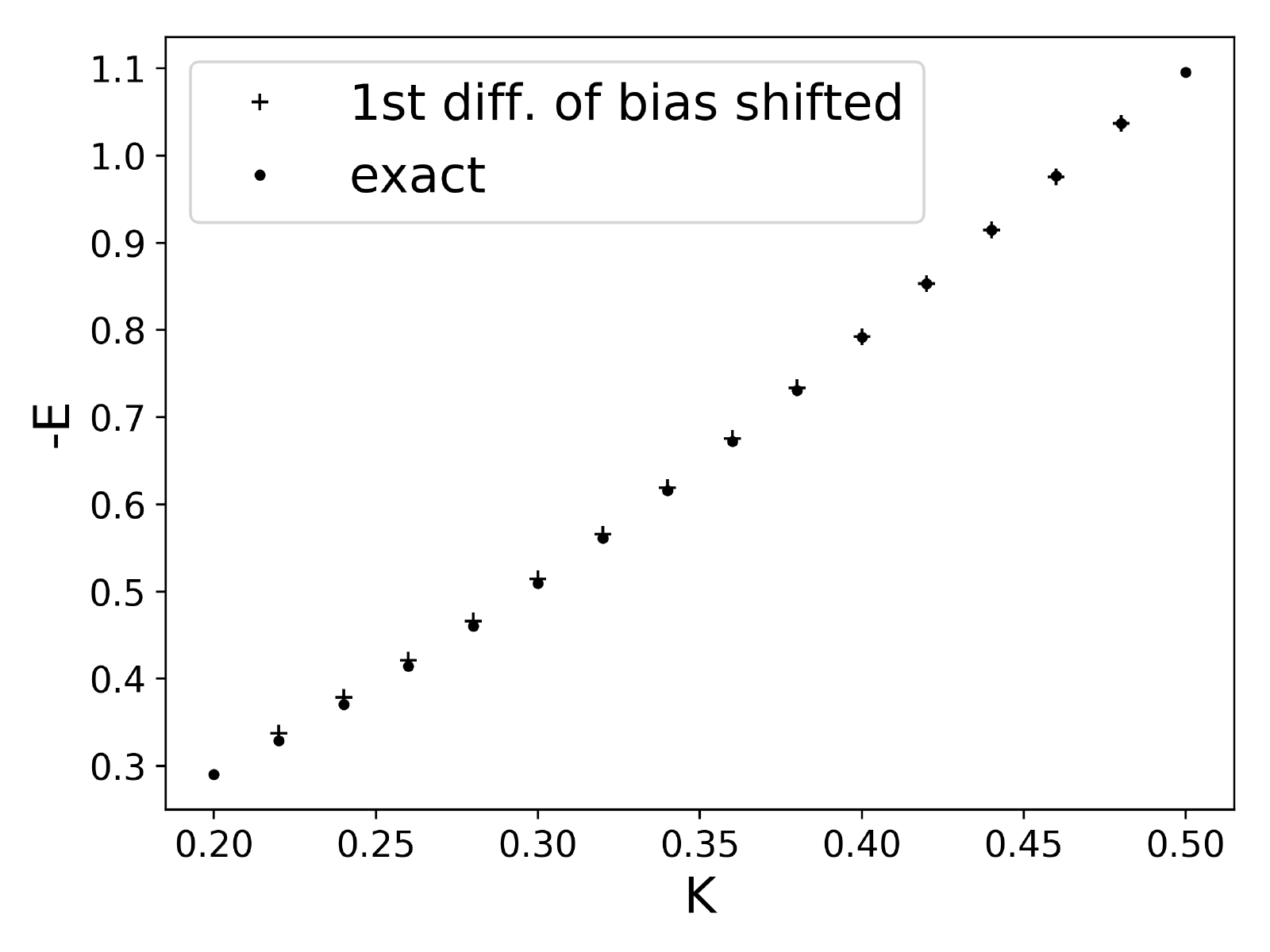}
\caption{First difference of $b_j$ shifted for comparison.}
\label{bias1Kshifted}
\end{center}
\end{figure}

\begin{figure}[!h]
\begin{center}
\includegraphics[clip,scale=0.6]{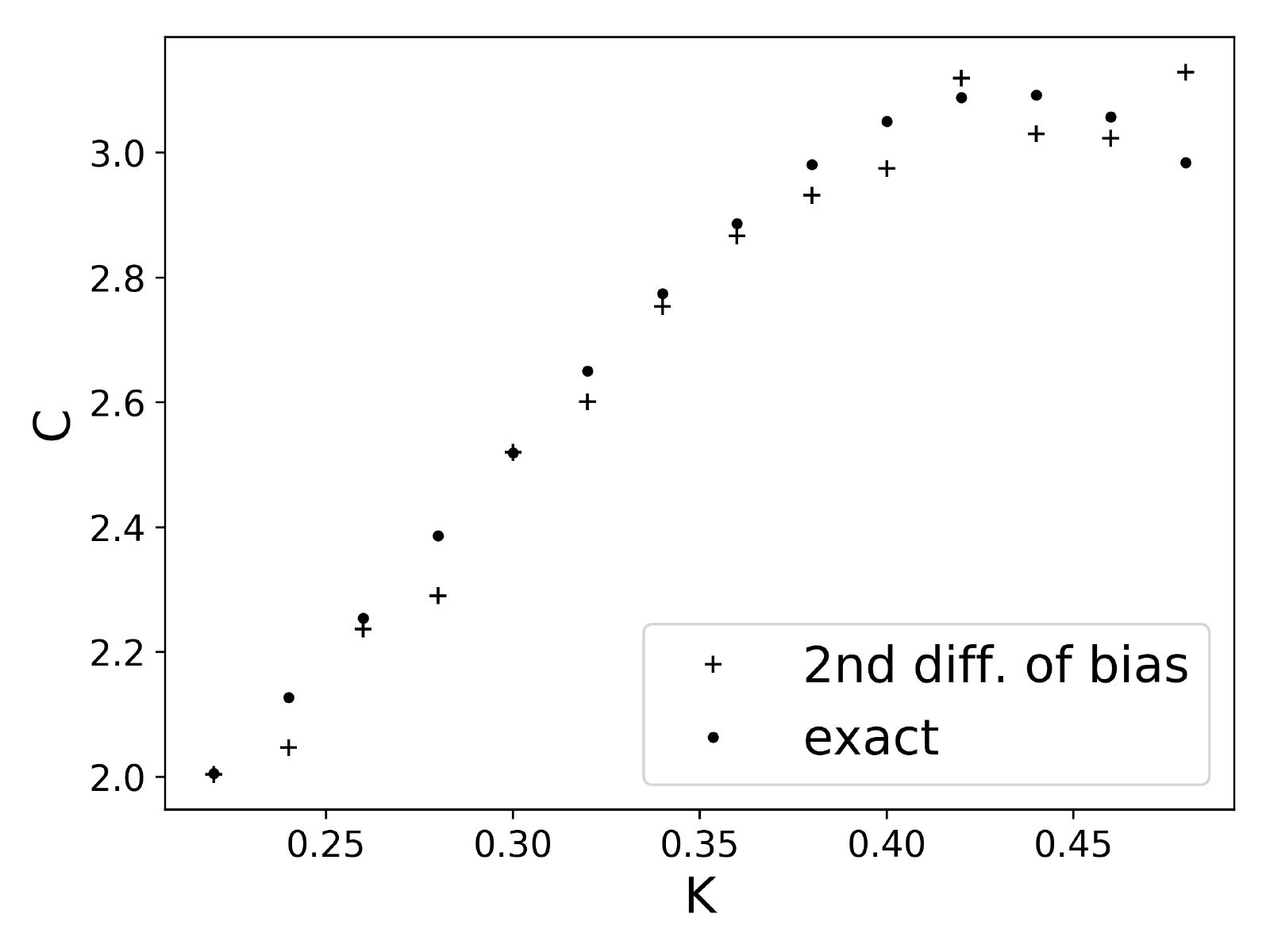}
\caption{Second difference of $b_j$ and exact specific heat.}
\label{bias2K}
\end{center}
\end{figure}

\begin{figure}[!h]
\begin{center}
\includegraphics[clip,scale=0.6]{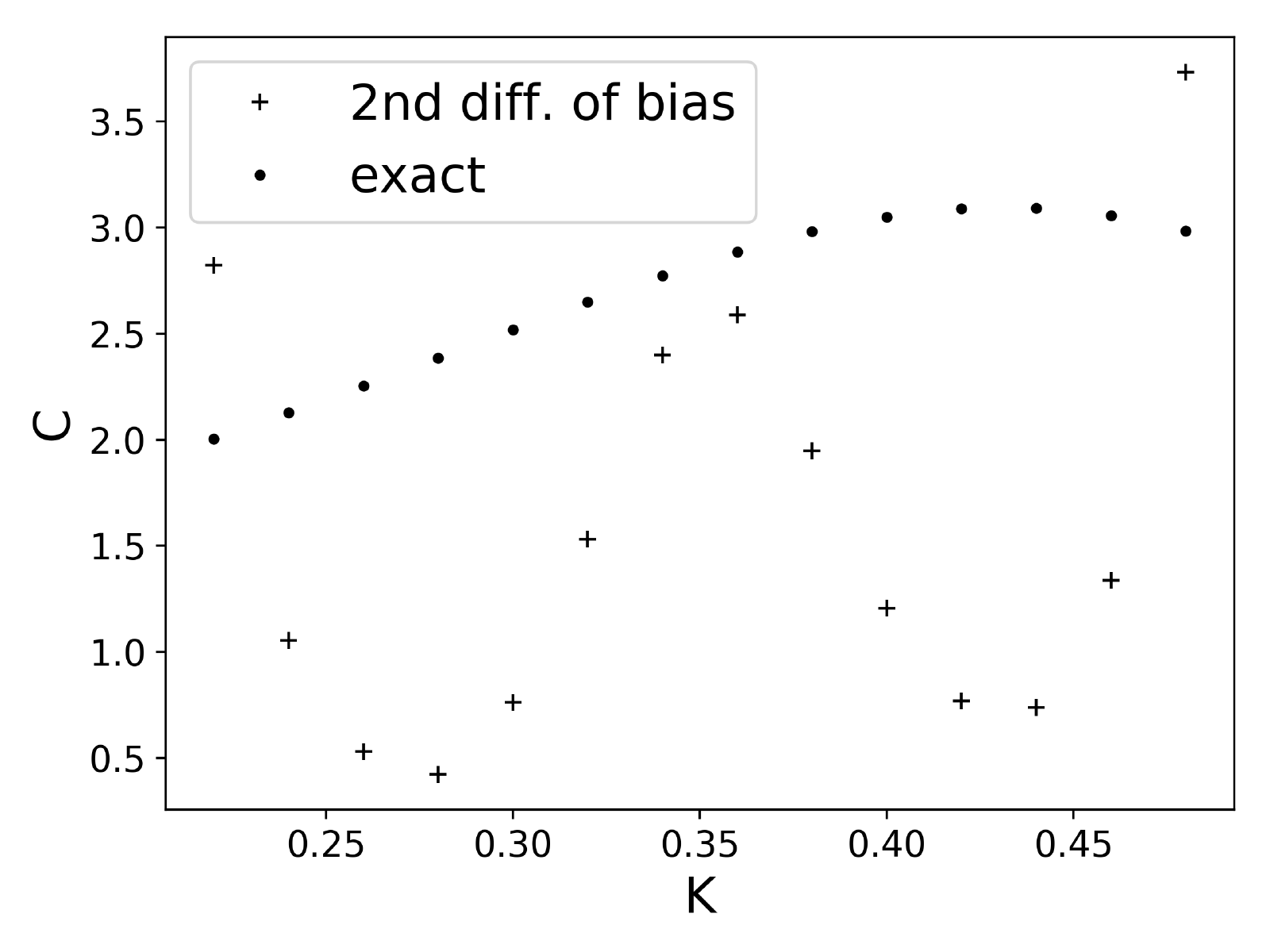}
\caption{Second difference of $b_j$ for $N_{\rm a}=4$ learning.}
\label{bias2KNa4}
\end{center}
\end{figure}

We calculate the first difference of the optimized bias according
to Eq.~(\ref{biasfirstdifference}) to evaluate the energy expectation 
value as a function of temperature.
In Fig.~\ref{bias1K}, we plot the first difference and the exact
values calculated by the BDRG. 
As is explained in Sec.~4, we see a constant
shift between these two plots due to the arbitrariness of the origin of
the energy. Apart from this constant shift, the coincidence 
is excellent, which is demonstrated in Fig.~\ref{bias1Kshifted},
where we shift the plot to coincide at $K=0.44$.

Then, we proceed to the second difference of the bias
defined in Eq.~(\ref{biasseconddifference}) and plot it
in Fig.~\ref{bias2K}. We also plot the exact value
of this finite system calculated by BDRG.
Although some statistical fluctuations remain,
we can clearly see that the second difference of the bias 
reproduces the specific heat well quantitatively and
the peak singularity appears, which 
is the phase transition remnant.
Note that the rightmost point does not appear good, 
the reason for which must be the edge effect 
in the temperature selection.
This type of edge anomaly can also be seen in 
Figs.~\ref{entropydensity} and \ref{HRR1}.

Here, we add another result using the bundled configuration
method introduced in Eq.~(\ref{bundledconfiguration}), where
we have a higher accuracy limit. The optimized machine
actually achieves a higher accuracy result, $0.702$, but
it is somewhat low compared with the theoretical limit, $0.72154$, 
even after 2.9M learning cycles.
Also the error function still remains large, $0.720$, compared with
the lowest limit, $0.640$.

Thus, the bundled configuration learning is not effective
and is not well tuned; rather, the situation is 
definitely worse than the single-configuration learning. 
We show the second difference of the bias in Fig.~\ref{bias2KNa4}.
The shape of the plot is far from the exact values, which is understandable
from the error function value.

Of course, also for the bundled configuration case, the perfectly
optimized machine should obey our solution as well.
However, owing to the higher accuracy limit, 
the requirement in the machine tuning is obscurer 
than that in the single-configuration case, and
conversely, the optimization procedure becomes subtler and harder,
and the final physical output level decreases. 
This result is very interesting. We recall the saying that
loose education does not make a person.

\begin{figure}[!h]
\begin{center}
\includegraphics[clip,scale=0.6]{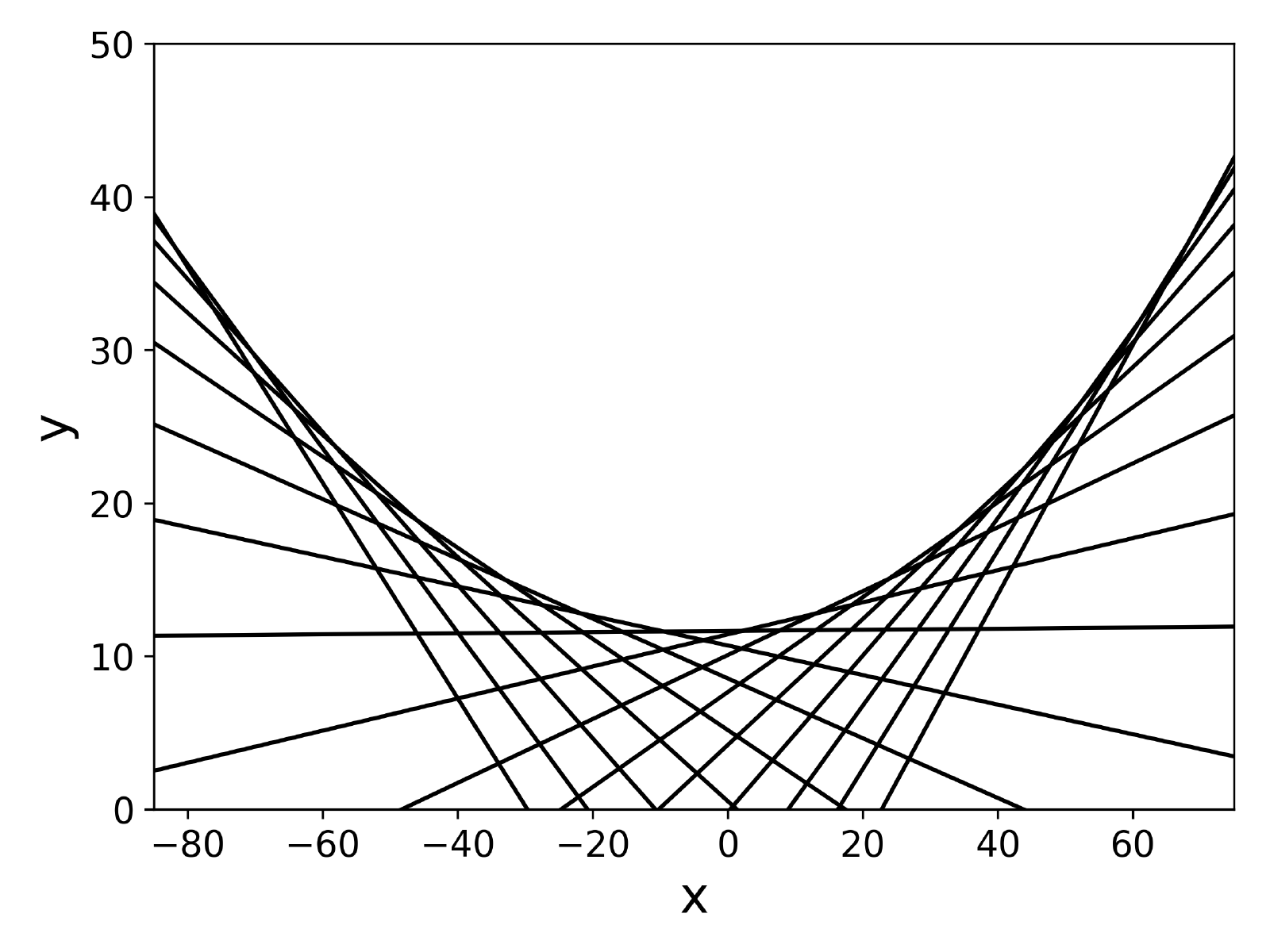}
\caption{Envelope of $y_j$ functions.
from $K=0.2$ (right) to $K=0.5$ (left) in order.}
\label{yfunction}
\end{center}
\end{figure}

Now, we look at another important functions defined in 
the full connection layer,
\Beq
y_j = w_j x + b_j.
\Eeq
These linear functions of $x$ will be the input of the softmax
function to make the final output.
We plot these functions in Fig.~\ref{yfunction}.
The variable $x$ 
represents the energy of the input configuration.
We select the highest $y_j$ for each $x$, which will be the
maximum likelihood estimates of the temperature.
We see contiguous changes of the output
temperature in Fig.~\ref{yfunction}.

Now, we can understand important facts about the difference
between the maximum accuracy and the minimum error conditions.
The maximum accuracy can be achieved if the output
temperature class obeys the step function depicted in
Fig.~\ref{ene-cla1}. In Fig.~\ref{yfunction}, this condition is seen
as follows. The crossing points of the first and second 
upper lines represent
the stepping points of energy. Keeping the
$x$ coordinate of these points the same,
we can move these $y_j$ lines without decreasing the
total accuracy. 
What the highest line is does matter, but the positions of other
lines below it do not matter. 
Since we have many (17 in this case) degrees of freedom
to move, there are many completely flat directions 
of the total accuracy in the parameter space of 
$w_j$ and $b_j$.

On the other hand, we have only four arbitrary parameters
for the flat directions giving the minimum error.
This is because the minimum error does require 
complete coincidence of all output temperature posterior
probabilities at any energy. Therefore, the positions of all
lines at any energy do matter to ensure the 
minimum error.
Therefore the minimum error condition is very much
stronger than the maximum accuracy condition. 

In Fig.~\ref{yfunction}, we also notice that an 
envelope function of these $y_j$ lines appears.
As obtained in Eq.~(\ref{basicequality}), 
the optimized machine parameter should obey
the equality
\Beq
y_j=w_j  x + b_j = -K_j E + F_j
\Eeq 
up to a linear function in $E$.
At each point on the envelope, the value
$E$ is simply the energy expectation value 
at the dominant temperature, $E^*_j$.
Therefore, on the envelope we can write 
\Beq
y_j = - K_j E^*_j+ F_j = S_j.
\Eeq
Thus, the envelope curve seen in Fig.\ref{yfunction} 
represents the entropy 
of the statistical system as a function of energy
up to a linear function in $E$.
These arbitrary parts do not contribute when we
take the second derivative of the envelope curve,
which also gives the specific heat again.

\section{Concluding Remarks}

We analyzed a simple model plant where a statistical
system with a definite temperature generates configurations, 
which are transfered to the input data of a deep learning machine for the 
supervised learning of the statistical system temperature.
We prepared 16 temperature classes. The temperature value itself
does not matter at all here. Actually, we have no way of determining the
temperature value of a physically realistic 
system without a verified thermometer.
That is, we do not control the temperature value itself, but we just 
set up some equilibrium states with different temperatures by 
adjusting the ``heating'' power for the pan.

Thus, our setup of the system is totally realistic, that is, we have
a statistical system that may realize a number of equilibrium 
states with different temperatures. There is no need for knowledge of
the temperature values themselves. Even in this situation, 
the deep learning machine can detect the specific temperature 
class that corresponds to the possible phase transition point.

To evaluate objectively the level of learning or perfectness of the machine, 
we theoretically calculate the upper bound of the accuracy and the
lower bound of the error function (cross entropy) in advance for the
input configurations.
The optimized machine actually gives scores that are very near 
to these theoretical bounds.

As proved in Sec.~2, 
the optimized machine must know the Hamiltonian of each 
configuration to realize the theoretical bound. 
Our machine is organized so that the input data are 
transformed by multilayered convolutional filtering and
by averaging the space direction into the single
variable $x$. Then, this variable must be the Hamiltonian 
up to the origin and the unit of energy.
We confirmed this property of the simple correspondence
between the key variable $x$ and the energy of the input
configuration as drawn in Fig.~\ref{energyvsx}.

Then, the final section of the machine is a full connection
layer connecting the variable $x$ and 
the output temperature classes through 
the machine parameters: weight $w_j$ and bias $b_j$,
where $j$ labels the output temperature class.
The condition of minimizing the error function defined 
by the cross entropy is satisfied by the linear
equations for $w_j$ and $b_j$ obtained in Eqs.~(\ref{optweightfunction}) and 
(\ref{optbiasfunction}).

Now, the temperature values can be read off through the weights $w_j$
up to the unit scaling and the origin,  and as a result, $K_j$ is obtained as 
a linear function of $w_j$. 
Then the bias $b_j$ is the free energy up to a linear function of
$K_j$. This is a remarkable result. The difference, a linear function, remains
undetermined, 
but this is natural since it comes from the freedom of the 
total normalization of the partition function $Z$ and the origin 
of the energy. These two quantities are irrelevant here, or impossible to
determine by our machine learning
method of analyzing the probability density of the statistical system. 
Then, the second derivative of free energy, the specific heat or
the fluctuation of energy, is obtained without any arbitrary parameter.

The free energy is a physical quantity that has more 
detailed information of the statistical system than the energy itself.
Its evaluation needs the information of entropy, the number of 
independent states with the energy fixed.
Therefore, it appears somewhat miraculous that the deep learning machine
engraves the free energy on its optimized bias parameters.

The key logic is that the optimization is performed 
to make the probability density of the energy of the machine output
equal to that of the input mixed temperature set of the configurations. 
To evaluate the probability density, we need the entropy information
or the free energy in addition to the energy, since we have to 
normalize them properly for each temperature.
Also, the basic machine structure, that is, the key variable $x$ and the softmax
function connecting it to the output temperature classes, 
ensures that the optimization condition can be actually satisfied,
which finally records the free energy on the bias parameters.

Here, we should stress again that the above remarkable results come
from the optimization condition that we adopted.
Our machine is trained to minimize the cross entropy.
However, the minimum cross entropy is not a necessary condition
for the maximum accuracy, although it is a sufficient condition.

The reason for using the cross entropy instead of the 
accuracy to tune the machine parameter is simply that the
stochastic steepest descent method cannot be applied to maximize 
the accuracy. Minimizing the cross entropy can be dealt with  
straightforwardly by the stochastic steepest descent.
This optimization condition of the minimum cross entropy
allows us to deduce our conclusion about the optimized weights and 
biases of the full connection section.

For the input statistical system, 
we adopt the one-dimensional Ising model with long-range interactions and
two-dimensional Ising model with the nearest-neighbor interactions.
In both models, our results of the free energy extracted from the bias
parameters coincide well with the exact calculations or MC simulation results.
The peak structures correctly indicate the remnant of the phase transition
in both models.

Our results show that the machine trained to guess the temperature
must recognize the Hamiltonian of the system, which is the conjugate
variable of the temperature in the probability density function.
In other words, by learning the temperature of configurations, the machine
actually becomes the spectrometer of the Hamiltonian, the conjugate variable
of the temperature.
As stressed when explaining Fig.~\ref{spectrometer}, 
the prediction of the temperature is just a way of learning, and
actually the accuracy of prediction cannot be better than 
the theoretical upper limit, which is a rather low value.
However, owing to the learning, 
the machine obtains the ability to calculate the
Hamiltonian, and this ability can possibly be perfect.

We should mention here 
that the optimized weights and biases 
are simply what is memorized in the neural network as a result
of learning.
Therefore, we understand the total procedure as follows.
By learning the temperature of each input configuration, 
the machine obtains the ability
to evaluate the energy (conjugate of the temperature), 
leaving the free energy and temperature values on the neural 
network couplings. This is the memory newly constructed
due to the learning in addition to the energy spectrometer 
function engraved on the lower layer parameters. 

We can generalize this structure for multiple control parameters, 
for example, the temperature and the external magnetic field.
Suppose we have $i_{\rm M}$ control parameters $K^{(i)}$ and 
the probability density is written as 
\begin{equation}
P[\Haii] = \frac{\exp(-\sum_i^{i_{\rm M}} K^{(i)}H_i[\Haii])}
{Z(K^{(i)})},
\end{equation}
where $H_i[\Haii]$ is the corresponding Hamiltonian.
Note that control parameters can be understood also as
individual interaction coefficients, as long as they can be
controlled from outside the statistical system.

We make configurations for a set of parameters $K^{(i)}$ and
transfer them to the input configuration of the deep learning.
We denote the $K^{(i)}$ value of set (class) $j$ by $K_j^{(i)}$.
To achieve the highest possible scores of accuracy, 
the deep learning machine must know all $i_{\rm M}$ functions
$H_i[\Haii]$, which are simply the conjugate variables of
$K^{(i)}$.
After some convolutional layers and averaging in the space direction,
we must keep $i_{\rm M}$ key variables, 
$x_i$, which represent
$H_i[\Haii]$. The final full connection layer is defined
by weight $w_{ji}$ and bias $b_j$ as follows:
\begin{equation}
y_j = \sum_i^{i_{\rm M}} w_{ji} x_i + b_j.
\end{equation}
As before, the softmax function outputs the probability
\begin{equation}
q_j =  \frac{e^{y_j}}{\Sigma_k e^{y_k}}.
\end{equation}

Following the arguments in Sec.~4, 
we reach the conclusion that the optimized
parameters must obey the analog of 
Eqs.~(\ref{optweightfunction}) and (\ref{optbiasfunction}), 
\begin{eqnarray}
w_{ji}&=& a_{1i} K_j^{(i)} + a_{0i}, \\
b_j&=&F(K_j^{(i)}) + \sum_i^{i_{\rm M}} c_{1i} K_j^{(i)} + c_0,
\end{eqnarray}
where $a_{1i}, a_0, c_{1i},$ and $c_0$ are arbitrary constants.
Thus, the optimized machine parameters have the 
same characteristic structures:
the weight $w_{ji}$ obeys a linear function of $K_j^{(i)}$
in the $j$ direction. 
The second derivative of the free energy 
with respect to $K^{(i)}$,
\begin{equation}
\frac{\partial^2F}{\partial K^{(i)} \partial K^{(k)}},
\end{equation}
can be evaluated without ambiguity.

We did not argue about the machine structure in detail.
Actually, the simple standard form of deep learning with
multilayered convolutional filtering 
worked well and achieved scores very near the theoretical bounds.
However, it is still not perfect of course. 
For one-dimensional nearest-neighbor 
interaction case, we can easily make the perfect machine within
this standard framework of the machine structure.
Although the possible perfect ability of the machine to approximate
any function has been argued,\cite{irie, funahashi,Hornik,Cybenko,barron} 
the actual machine with a finite 
number of components has limited performance.
As a result, we should say that our simple machine fortunately 
obtains a sufficiently high power 
so that we can observe the phase transition remnant peak of
the specific heat 
in the second difference of the bias parameters.

We guess that the reason why such convolutional filtering
succeeds in calculating the Hamiltonian is that
the Hamiltonian constitutes the relatively local spin interactions.
In this point, the renormalization group method of summing up
local fluctuations to define effective macro-interactions should
shed light on the effectiveness of the convolutional layers.
However, the detailed and precise correspondence between 
the renormalized variables and the variables on intermediate
layers has not yet been clarified.
It should also be interesting to test the Hamiltonian recognition ability
for various Hamiltonians including unphysical
and unnatural interactions, such as long-range dominated ones.

We proved that the optimized machine must know the 
system Hamiltonian so that it may achieve the highest accuracy.
Then, one may claim that such a machine 
can be trained more directly by making a function-approximating 
machine with supervised learning.
However, note that for such supervised
learning, we need the exact Hamiltonian function to label every
input configuration.
This is not usually the case for realistic issues.
Our input label is just the temperature class used to generate 
the configuration. This is sufficient for the
machine to learn the Hamiltonian and to engrave the free energy 
of the system.

The original motivation of our research was to clarify the
relationship between the deep learning and the renormalization
group. However, now, we notice that there exists rather large 
differences between the deep learning machine 
to learn the temperature
and the renormalization group analysis, although their networking
structures resemble each other closely.

The temperature is the
target of deep learning, and the machine handles configurations
generated by all classes of temperature simultaneously.
On the other hand, for the renormalization group procedure, the 
temperature is the coefficient of the most relevant
operators controlling the system. Strictly speaking, the
relevant operator contains the Hamiltonian as its component
and the temperature may play a role of labeling the
size of the relevant operator.

The renormalization group describes the change of this
label, the temperature parameter, as the lower layer
variables are renormalized into upper layer variables.
That is, the temperature changes from lower to upper
layers, and it is enlarged, which means that it is relevant, 
while other operators are reduced to vanish and are
thus called irrelevant.

We should note another apparent difference between the 
deep learning machine and the renormalization group.
The weights and biases defining the filters in the 
convolutional network are optimized independently for
each layer. These parameters define the upper layer
variables in terms of the lower layer variables, and
are simply the definition of the renormalization.

The renormalization group transformation
is defined commonly for all layers. This is an essential feature
of the renormalization group, that is, applying the same transformation
many times. Then, after identifying the 
fixed point of the transformation, which corresponds to the criticality
of the system, we linearize the
renormalization group transformation in the neighborhood of the fixed point. 
Then, the transformation is characterized by its eigenvalues. 
Repeating the same transformation results in 
the enhancement and reduction of operators
depending on the absolute value of their eigenvalues, 
which are greater than unity (relevant) or less than unity (irrelevant).

Thus, changes of the operators are described by a geometric series near 
the criticality. The relevant operator changes with 
a ratio greater than unity, and the ratio can be observed as the critical 
exponent through the experimental observation around criticality.
This is simply the essential framework for understanding
the second-order phase transition by using the renormalization 
group.
This basic logic of the renormalization group analysis
looks considerably different from the optimized machine
parameter structures.
However, note that this difference should not be understood
as a contradiction.
The optimization of the renormalization transformation independently for
each step may give a new direction for the development of the renormalization 
group method.

Keeping these points in mind, we consider that the restricted
Boltzmann machine (RBM) is expected to have tighter correspondence 
with the renormalization group.\cite{KT}
The lower-layer visible variables 
are the microvariables whose probability distribution
is given by the micro Hamiltonian, and the upper-layer hidden 
variables are regarded as the renormalized macrovariables.
The weights and biases defining the RBM are simply the
rules of defining the renormalization of microvariables to
give the macrovariables.
The upper layer macro-variables should follow the renormalized Hamiltonian.
These layers can be stacked deeply similarly to the recursive
renormalization group transformations.

We consider an RBM and we use the visible variables as the input
configuration of the deep learning machine.
We do not know the Hamiltonian controlling the visible
variables. Then, can we obtain information about the Hamiltonian
through the optimized deep learning machine parameters?

To learn something about the Hamiltonian, we need a set of 
different values of the conjugate of the Hamiltonian, 
the temperature. However, there is probably no simple way of adjusting the
temperature with a fixed Hamiltonian by changing the 
RBM weight and bias parameters.
The number of degrees of freedom of the Hamiltonian controlling
the RBM visible variables is much larger than that of
RBM machine parameters. Note that the RBM-defined Hamiltonian
contains multispin interactions in addition to the basic 
two-spin interactions, and thus the possible number of 
different sorts of interactions is quite large.
Therefore, we need more detailed analysis of the RBM and
its effective Hamiltonians before moving forward.

Anyway, we hope that the in-depth mechanism of
the high performance of the deep learning machine should 
be related to the mechanism of the renormalization 
group to treat many degrees of freedom by coarse graining
and to pick up the characteristic variables
automatically as relevant operators.
On the other hand, we also hope that the future understanding of 
the origin of deep learning machine power will
give us new ideas for developing the renormalization group method, 
which is already appreciated as a universal tool 
in various fields.

\vskip10mm
\noindent
{\bf Acknowledgments}

We acknowledge fruitful and helpful suggestions and 
discussions with 
Yasuhiro Fujii and 
Shin-Ichiro Kumamoto.
We appreciate collaborative works of ERBM by
Mariko Iijima and of MC simulation by Tsubasa Jingu, and 
the initial setup of our computing system and
preparation studies in the early days of this project by
Yuusuke Hori and Shinnosuke Onai.
Also, many thanks should be given to Muneki Yasuda for his excellent lectures and 
for motivating us to study the relationship between
the deep learning and the renormalization group.

\end{document}